\documentclass[usenatbib]{mnras2}

\usepackage[T1]{fontenc}

\DeclareRobustCommand{\VAN}[3]{#2}
\let\VANthebibliography\thebibliography
\def\thebibliography{\DeclareRobustCommand{\VAN}[3]{##3}\VANthebibliography}

\usepackage{graphicx}	
\usepackage{amsmath}	
\usepackage{amssymb}	
\usepackage{lipsum}
\usepackage{color}
\def\gtrsim{\lower.5ex\hbox{$\; \buildrel > \over \sim \;$}}
\newcommand{\msun}{\mbox{${\rm M}_{\odot}$}}

\usepackage{newtxtext,newtxmath}

\title[On the impact of runaway stars on dwarf galaxies]{On the impact of runaway stars on dwarf galaxies with resolved interstellar medium}
\author[U. P. Steinwandel et al.]{Ulrich P. Steinwandel,$^{1}$\thanks{E-mail: usteinwandel@flatironinstitute.org (UPS)}
Greg L. Bryan,$^{2,1}$
Rachel S. Somerville,$^{1}$
Christopher C. Hayward$^{1}$
\newauthor
and Blakesley Burkhart$^{1,3}$\\
$^{1}$Center for Computational Astrophysics, Flatiron Institute, 162 5th Avenue, New York, NY 10010\\
$^{2}$Department of Astronomy, Columbia University, 550 West 120th Street, New York, NY 10027, USA\\
$^{3}$Department of Physics and Astronomy, Rutgers University, 136 Frelinghuysen Road, Piscataway, NJ 08854, USA
}

\date{Accepted XXX. Received YYY; in original form ZZZ}

\pubyear{2022}

\begin{document}
\label{firstpage}
\pagerange{\pageref{firstpage}--\pageref{lastpage}}
\maketitle

\begin{abstract}
About ten to 20 percent of massive stars may be kicked out of their natal clusters before exploding as supernovae. These ``runaway stars'' might play a crucial role in driving galactic outflows and enriching the circumgalactic medium with metals. To study this effect, we carry out high resolution dwarf galaxy simulations that include velocity kicks to massive stars above 8 M$_{\odot}$. We consider two scenarios, one that adopts a power law velocity distribution for kick velocities, resulting in more stars with high velocity kicks, and a more moderate scenario with a Maxwellian velocity distribution. We explicitly resolve the multi-phase interstellar medium (ISM), and include non-equilibrium cooling and chemistry. We adopt a resolved feedback scheme (\textsc{Griffin}) where we sample individual massive stars from an IMF and follow their radiation input as well as their the SN-feedback (core-collapse) channel at the end of their lifetime. In the simulations with runaway stars, we add additional (natal) velocity kicks that mimic two and three body interactions that cannot be fully resolved in our simulations. We find that the inclusion of runaway or walkaway star scenarios has an impact on mass, metal, momentum and energy outflows as well as the respective loading factors. We find an increase in mass, metal and momentum loading by a factor of 2-3, whereas we find an increase in the mean energy loading by a factor of 5 in the runaway case and a factor of 3 in the walkaway case.  We conclude that the inclusion of runaway stars could have a significant impact on the global outflow properties of dwarf galaxies.
\end{abstract}




\section{Introduction}
The $\Lambda$CDM paradigm has been extremely successful at predicting and reproducing observations of structure formation on large scales (greater than a few Mpc). However, since the first attempts to model galaxy formation within this paradigm, it has been clear that \emph{feedback} from massive stars, supernovae, and active black holes is critical for reproducing many observed properties of galaxies within this paradigm \citep[e.g][for reviews]{Somerville2015, Naab2017}. There has been great progress recently in developing large scale simulations that reproduce many key observations of galaxy populations \citep[e.g.][]{Hirschmann2014, Schaye2015, Vogelsberger2014, Pillepich2018, Nelson2018}. In addition, in the past decade, numerical simulations finally succeeded in producing thin, rotation dominated disks with rather inefficient baryon-to-star conversion \citep[e.g.][]{Guedes2011, Agertz2013, Marinacci2014, Hopkins2014, Hopkins2018}. 

However, all of these simulations to one degree or another achieve this success by implementing somewhat ad hoc or phenomenological ``sub-grid'' recipes, as they are not able to explicitly resolve many of the physical processes that actually drive feedback. For example, large-volume simulations typically have a spatial resolution of a few hundred parsecs or larger, and baryonic mass resolution of $10^5 \msun$ or greater (see figure 1 of \citealp{nelson2019} for a recent summmary). Thus these simulations typically do not resolve the multi-phase interstellar medium (ISM), nor blastwaves from (individual) supernova explosions, among many other relevant processes. One of the biggest open questions in galaxy formation theory currently is how ``small scale'' processes (that occur below the explicit resolution of the simulation) may impact larger scale and \emph{global} properties of galaxies. 

A new generation of simulations are achieving parsec to sub-parsec spatial resolution and mass resolution comparable to that of single stars. These simulations can explicitly resolve the multiphase ISM, the Sedov-Taylor phase of individual supernova explosions, and the multiphase character of outflows, and are therefore able to implement the physics of star formation and stellar feedback in a more explicit manner. Several groups have carried out such simulations for sub-galactic regions \citep{Kim2015,Kim2017,Kim2020,Walch2015}, typically representative of Solar Neighborhood or Milky Way conditions. Recently, a few groups have been able to achieve this ``individual star'' resolution and implement these more explicit feedback approaches within simulations of \emph{entire dwarf galaxies} with accurate non-equilibrium heating and cooling \citep{Hu2017,Emerick2019,Lahen2020a,Smith2021}, which is a significant step forward. In this work, we leverage this relatively new capability to study a process that is unresolved in galaxy scale simulations, but may play an important role in shaping ``macroscopic'' galaxy properties: runaway stars. 

Massive stars are predominantly born in clusters. Simulations of individual star clusters \citep[e.g.][]{Oh2016} have shown that some fraction of massive stars can receive kicks large enough to move them out of their natal clouds before exploding as supernovae. These kicks can be imparted by gravitational interactions within a binary star system (sometimes involving a third star), or when one of the stars in a binary explodes as a SNae. In both cases, one of the massive stars can be accelerated above the escape velocity of the binary system (or even the star cluster itself). These unbound stars can easily travel at velocities $> 30$ km s$^{-1}$ and can thus travel the typical length scale of a GMC of a few $10$ parsec on a timescale of $~1-2$ Myr. However, we note that it is not clear yet which of the scenarios for the formation of runaway stars is favoured \citep{Renzo2019, Fujii2011}. 

It is extremely challenging to include these processes in a galaxy scale simulation, for several reasons. First, one would have to model the binary stellar intial mass function (IMF), which has significant uncertainties \citep[e.g.][]{Kroupa1995, Malkov2001}. Second, and perhaps more challenging, one would have to model two and three body interactions of single stars. Galaxy formation simulations are not well equipped to carry out orbital integration of multiple stellar systems, due to the nature of the gravity calculation which is often performed with a tree or particle mesh. Hence, the gravity solvers of many astrophysical simulation codes operate in the limit of a \textit{collisionless} system, while this is no longer a good description of few body interactions in star clusters. At the cloud scale, there are attempts to circumvent this weakness \citep[e.g.][]{Hirai2021, Fujii2021, Grudic2022, Guszejnov2022}. However, it has also been pointed out recently that there might be an issue with the underlying Schmidt-type star formation recipes often applied in galaxy scale simulations which makes it intrinsically hard to disperse star clusters on the correct timescales \citep[e.g.][]{Smith2021, Hislop2022}, a problem which would likely become worse when orbits are integrated using direct summation algorithms. 

Why might we expect that runaway stars could have a significant effect on galaxy scale properties? Supernova (SNe) feedback is believed to have the strongest impact on providing the necessary mid-plane pressure to drive a galactic outflow. \citep{Martizzi2015, Kim2015, Walch2015, Gatto2017, Naab2017, Steinwandel2020}. However, recently several groups have put forward the idea of galactic winds that are driven by cosmic-rays \citep[e.g.][]{Buck2020, Girichidis2016, Girichidis2022, Hopkins2021d, Hopkins2021e, Hopkins2021a, Hopkins2021b, Hopkins2021c}. For SN-feedback, the crucial condition to drive outflows is the formation of a hot volume filling ISM phase \citep[e.g.][]{Naab2017, Steinwandel2020}. An important factor in this picture is the host environment of the SNe \citep[e.g.][]{Hu2016,Hu2017, Hu2019, Lahen2020b, Lahen2020a, Gutcke2021, Smith2021, Hislop2022}. SNe that explode in the cold neutral medium (CNM) or the cold molecular medium are inefficient in establishing the hot phase of the ISM as cooling times are short (a few 100 years) and the remnants remain small ($\sim10$ pc) while in the energy conserving Sedov-Taylor phase. This makes it relatively unlikely that subsequent supernovae will explode in the bubble of previous SNe, so that no efficient outflows can be launched from the cold ISM due to the formation of superbubbles \citep[e.g.][]{Fielding2017,Orr2022a,Orr2022b}. However, the situation is different in the diffuse ISM at lower densities and higher temperatures, where SNe remnants occupy a larger volume fraction. Cooling times are much longer, and a hot phase is naturally established, due to the larger occupation fraction of the remnants which makes superbubble formation and subsequent breakout much easier. \\
There are a limited number of physical processes that can lead to lower ambient media preceding a SN-event, of which some are driven by the host star itself \citep[e.g. ][for an overview of main and post main sequence stellar winds, photo-ionisation, photoelectric heating]{Forbes2016, Haid2016, Hu2017, Haid2018, Emerick2019, Kim2019, Smith2021, Lancaster2021a, Lancaster2021b, Lancaster2021c}. As discussed above, runaway stars are another mechanism that could potentially allow massive stars to travel out of their dense natal gas clouds, such that the SNe potentially explode in a warmer, lower density environment, enhancing their ability to drive large scale outflows \citep[e.g.][]{Naab2017}.

The impact of runaway stars has recently been studied in conditions similar to the Milky Way in \citet{Kim2018} and \citet{Andersson2020}. However, there has been no detailed study of this issue in the context of dwarf galaxies, which is the goal of this paper. In this paper, we create 4 simulations of isolated dwarf galaxies similar to the observed Wolff-Lundmark-Moyette (WLM) system at the outer edge of the Local Group (baryon mass of $2 \cdot 10^7$ M$_{\odot}$ and dark matter mass of $2 \cdot 10^{10}$ M$_{\odot}$). We use the  \textsc{Griffin}(Galaxy Realizations with the feedback from individual massive stars) model for single star formation and resolved feedback \citep[e.g.][]{Lahen2020b, Lahen2020a, Steinwandel2020, Hislop2022}. We include a sub-grid treatment for runaway stars based on existing results for the runaway population from simulations and observations \citep[e.g.][]{Eldridge2011, Oh2016}. We note that our approach includes only the natal kicks, neglecting the gravitational interaction of the star with the environment over its lifetime, and therefore may represent an upper limit on the effects of including such stars in galaxy simulations. \\
The remainder of the paper is structured as follows: In Section \ref{sec:numerical} we introduce the code, present the \textsc{Griffin} model for single star formation and resolved feedback, and explain in detail how we incorporate runaway stars into our model. In Section \ref{sec:results} we discuss the results with a specific focus on how runaway stars may alter outflow properties of dwarf galaxy systems and impact the ISM and CGM-ISM interface. In Section \ref{sec:discussion} we discuss our results and compare to previous simulations in the literature. In Section \ref{sec:conclusion} we summarise our results.

\section{Numerical Methods}
\label{sec:numerical}
In this section we discuss the details of our numerical modelling and present the initial conditions of our simulations. The results are part of the \textsc{Griffin} project \footnote{\url{https://wwwmpa.mpa-garching.mpg.de/~naab/griffin-project/}}.

\subsection{Simulation Code}
For all simulations presented we use the Tree-SPH code P-Gadget3 \citep[][]{springel05, Hu2014} to carry out the gravity and hydrodynamics calculations. However, we made significant improvements by updating the baseline SPH solver to incorporate the meshless finite mass (MFM) method for solving the fluid flow following the methods discussed in \citet{Hopkins2015} and \citet{Gaburov2011}, using the MFM implementation of \citet{Steinwandel2020}. MFM represents a fully consistent second order method in space and time and we compute fluid fluxes by solving the one-dimensional Riemann-problem between single fluid tracers by adopting an HLLC-Riemann solver that accounts for the appropriate reconstruction of the contact wave. The face area is defined over the smoothing-length weighted quadrature point between the single fluid tracers (Steinwandel in prep.).\\
The code is then coupled to the cooling and feedback network, utilised in the \textsc{Griffin}-framework \citep[e.g.][]{Hu2016, Hu2017, Hu2019, Lahen2020a, Lahen2020b, Steinwandel2020, Hislop2022}. 

\subsection{Cooling, heating and star formation}
To model the multiphase ISM, we self-consistently model the non-equilibrium formation and destruction of molecular hydrogen and CO, following the cooling and chemistry networks developed in \citet{Glover2007a} and \citet{Glover2012}, where we also account for the self-shielding of molecular gas based on a column density computation on a healpix sphere. In earlier versions of our code this was achieved by the \textsc{TreeCol} method. In the latest version this is done with a healpix expansion using the c library \textsc{chealpix}. The details of the cooling and chemistry network can be found in \citet{Hu2017} and \citet{Steinwandel2020}.\\ 
Star formation is modelled by utilising an IMF-sampling approach where we sample all the stars more massive than 4 M$_{\odot}$ (base resolution of the simulation) as single stars. The details of the method can be found in \citet[e.g.][]{Hu2017}; we have adopted minor updates here that will be presented in greater detail in Steinwandel (in prep.).

\subsection{Stellar feedback}
We follow several feedback channels for the massive stars that form within the simulation, including the photoelectric heating, the UV-radiation and the photoionising radiation, as well as the metal enrichment from AGB and subgrid stellar winds. However, the core of the stellar feedback scheme is the resolved SN-feedback mechanism developed by \citet{Hu2017} and \citet{Steinwandel2020}, that allows for a self consistent build-up of the hot phase and the momentum of individual SN-events in a resolved Sedov-Taylor phase \citep[see][]{Hu2021, Hu2022}. Hence, all the outflow properties in these simulations are self-consistently driven from the hot phase of the ISM without the need for any ``ad-hoc'' tuning of mass loading factors as adopted in lower resolution galaxy formation simulations. Therefore, our simulations allow for a detailed study of the origin and driving of large scale galactic winds. \\

We include the UV radiation of single stars following the implementation of \citet{Hu2017}. This implementation makes the assumption of a (locally) optically thin medium, which is a good approximation for dwarf galaxies. Hence, we can simply sum over the radiation field for each star using an inverse square law. The lifetime functions and effective surface temperatures are taken from \citet{Georgy2013}. The UV luminosity is obtained from the stellar library BaSel \citep{Lejeune1997, Lejeune1998, Westera2002} in the energy band $6 - 13.6$ eV, which dominates the photo-electric heating rate in the ISM.  We note that one star particle can contain more then one star in this respect as we trace the UV-radiation from all stars with masses larger than $2$ M$_{\odot}$. However, for star particles more massive than $4$ M$_{\odot}$ the stars fully represent single stars. We integrate the radiation contributions while walking the gravity tree. Hence, this is effectively a ray-tracing approach modulo the angular resolution of the tree. This can be straightforwardly improved in future implementations by either including higher order moments in the tree or modifying the tree's opening angle (i.e. running a separate gravity tree for the gravity force and UV-radiation computation) \citep[e.g.][for one approach to improve our method]{Grond2021}.\\
We follow photoelectric heating rates from \citet{Bakes1994, Wolfire2003} and \citet{Bergin2004} at the rate:
\begin{align}
    \Gamma_\mathrm{pe} = 1.3 \cdot 10^{-24} \epsilon D G_\mathrm{eff} n \ \ \mathrm{erg} \ \mathrm{s}^{-1} \mathrm{cm}^{-3},
\end{align}
with the hydrogen number density $n$, the effective attenuation radiation field $G_\mathrm{eff} = G_{0} \exp{-1.33 \cdot 10^{-21} D N_\mathrm{H, tot}}$, the dust-to-gas ratio D, the total hydrogen column density $N_\mathrm{H, tot}$ and the photoelectric heating rate $\epsilon$ that can be determined via:
\begin{align}
    \epsilon = \frac{0.049}{1 + 0.004 \psi^{0.73}} + \frac{0.037 (T/10000)^{0.7}}{1+2 \cdot10^{-4}\psi}
\end{align}
where $\psi$ is denoted as $\psi = G_\mathrm{eff} \sqrt{T} /n^{-}$, with the electron number density $n^{-}$ and the gas temperature $T$. 

We include a treatment for photoionising radiation since previous work \citep[e.g.][]{Hu2017, Emerick2019, Steinwandel2020, Smith2021}, has revealed that photoionisation feedback can have a significant impact on ISM properties and star formation rates. Ideally, we would utilize a proper radiative transfer scheme for this purpose, but this is prohibitively computationally expensive. Thus we model the photoionising feedback via a Stroemgren approximation locally around each photoionising source in the simulation, similar to the approach used in \citet{Hopkins2012} and later improved by \citet{Hu2017}. \\
Every star particle that fulfils the criterion of having one or more constituent stars with mass m$_\mathrm{IMF} > 8$ M$_{\odot}$ is treated as a photoionising source in our simulations. Thus all of the neighbouring gas particles are labelled as photoionised within the radius R$_{S}$ around the star and all of the surrounding molecular and neutral hydrogen is destroyed and transferred to H$^{+}$ (we set the ionising fraction to 0.9995 in our chemical model). Alongside this procedure we heat the surrounding gas within R$_{s}$ to 10$^4$ K, which is the temperature in HII-regions, by providing the necessary energy input from the ionising source. This procedure is straightforward if R$_{S}$ is known (i.e. as long as the gas density is more less constant). However, in practice this will rarely be the case, and we obtain $R_{S}$ iteratively, where we use the classic Stroemgren-radius as an ``educated guess''.\\
We include the feedback from Type II (CCSN) supernovae in our simulations. The supernova feedback routines presented in this section are the core feedback mechanism in our current galaxy formation and evolution framework. Every supernova distributes $10^{51}$ erg into the nearest 32 neighbours. We follow metal enrichment using the yields of \citet{Chieffi2004}, who present stellar yields for stars from zero metallicity to solar metallicity for progenitors ranging from 13 M$_{\odot}$ to 35 M$_{\odot}$, from which we interpolate our metal enrichment routines. When a star explodes in a CCSN the mass is added to the surrounding $32$ (one kernel) gas particles with the respective metal mixture that is expected from the above yields. It is important to include the correct metal yields in order to obtain the correct cooling properties of the ambient medium and thus in the end the correct density and phase structure of the ISM. However, as the amount of metal mass that is distributed into the neighbours of the star particle can be much larger than the mass of one gas particle, we split the gas particles after the enrichment has taken place if the mass exceeds the initial particle mass resolution of the simulation by a factor of two, to avoid numerical artefacts. The split particles inherit all the physical properties that were present in the parent particle including specific internal energy, kinetic energy, metallicity and chemical abundances. The particle mass is split by a factor of two and the position is offset by one fifth of the parent particle smoothing length in a random (isotropic) direction to avoid the overlap of the two newly spawned particles. The spawning process requires a new domain decomposition and a re-build of the gravity tree which makes the code somewhat slower.
\begin{table*}
    \caption{Overview of the physics variations adopted in our different simulations. For convenience we follow similar naming conventions as those adopted in \citet{Hu2017}.} 
    \centering
    \label{tab:models}
    \begin{tabular}{ccccccc}
      \hline
	  Name		& Core Collapse Supernovae & Photoionising radiation & Photoelectric heating & runaway stars\\
	  \hline
      \textit{WLM-fiducial} & \checkmark  & \checkmark & \checkmark & \text{\sffamily X}\\
      \textit{WLM-RunP} & \checkmark & \checkmark & \checkmark & power law distribution\\
      \textit{WLM-RunM} & \checkmark &  \checkmark & \checkmark & Maxwellian distribution\\
      \textit{WLM-inplane} & \checkmark & \checkmark & \checkmark & maxwellian (sampling only for v$_{x}$ and v$_{y}$)\\
      \hline
      \end{tabular}
\end{table*}
\subsection{Treatment of Runaway stars}
While the spatial and mass resolution of our simulation is relatively high by current standards, we are not yet capable of accurate modeling of two, three and few-body interactions needed to self-consistently account for the velocity offset from which runaway O/B stars originate. We note that attempts can be made to self-consistently model these phenomena in the near future in galaxy scale simulations employing the methods presented in \citet{Rantala2017} and \citet{Rantala2020} (higher order symplectic), and there are recent demonstrations of similar methods on the cloud scale for higher order hermite integration of star clusters \citep[e.g][]{Grudic2022, Guszejnov2022}. However, currently it is not possible to accurately compute the forces for two or three body encounters in a galaxy scale simulation that also takes a detailed ISM model into account. A first order approximation for the effect of runaway stars can be obtained by constructing a ``subgrid'' model for the velocity kicks for the massive O/B stars that become runaway stars. In the following we present two models, one motivated by simulations for a strong runaway case and another one which represents a weaker walkway case.

\subsubsection{The Runaway Case: Power law velocity distribution}
In the strong runaway model we assume that the velocity distribution follows a simple power law
\begin{align}
    f_\mathrm{vel} \propto v^{-\beta},
    \label{eq:distribution}
\end{align}
with the velocity power law slope $\beta$, which is dependent on several factors. One popular scenario for runaway star formation occurs when a binary system undergoes gravitational scattering processes with a third star. Thus it is straightforward to assume that $\beta$ is mainly driven by the number of binary systems in the star cluster and the relaxation time of the star cluster itself. Moreover, in reality, star clusters can be disrupted (e.g. due to internal supernova events or tidal interactions with other star clusters) which is also important for setting $\beta$. All these processes can lead to velocity kicks following the distribution of equation \ref{eq:distribution}. However, there would be a time delay that comes from the direct N-body interactions which is not captured by our approach. This is a clear limitation of applying random (isotropic) velocity kicks. 
To accurately model star cluster formation and evolution it is necessary to solve the direct N-body system, which would give a reliable estimate for $\beta$ \citep{Eldridge2011, Perets2012, Oh2016, Renzo2019}. In detail, the velocity distribution may depend on the mass of the star as well as the properties of the star cluster. Alternatively, one could follow an empirically motivated approach and adopt an estimate for $\beta$ directly from observations \citep[e.g.][]{Blaauw1961, Comeron2007, Gies1987, Hoogerwerf2000, Hoogerwerf2001}. 
In our first runaway model we will follow a theoretically motivated approach by adopting estimates from \citet{Oh2016}. This will allow for a direct comparison to \citet{Andersson2020}, who investigated the effects of runaway stars in galaxy simulations of the Milky Way.\\
We start our modelling by building a subgrid model for runaway stars in which we adopt a power law velocity distribution following eq. \ref{eq:distribution} with a slope $\beta=1.8$. To do so we slightly modify our IMF-sampling approach and add velocity kicks to stars that are more massive than $8M_{\odot}$. First, we have to sample the velocity distribution of equation \ref{eq:distribution}. This can be accomplished by drawing a random number $p$ and obtaining the absolute value for the velocity kick of particle $i$ in the following manner:
\begin{align}
    v_{i} = v_\mathrm{min} \cdot (1-p)^{-\frac{1}{\beta - 1}}.
    \label{eq:sampling}
\end{align}
We adopt a minimum value $v_\mathrm{min} = 3$ km s$^{-1}$ for direct comparison with \citet{Andersson2020}. We truncate the distribution at $385$ km s$^{-1}$. However, this truncation process introduces a complication. If we draw a random number between zero and one and we calculate $v_{i}$ following equation \ref{eq:sampling}, then it can happen that we naturally sample values above $385$ km s$^{-1}$. We want to truncate the sampling because this will give as a distribution where 14 percent of the stars obtain a velocity kick of more than $30$ km s$^{-1}$, which is in good agreement with observations. The sampling of values above $385$ km s$^{-1}$ can be related to a certain upper limit p$_\mathrm{max}$ for the random number $p$, which is given by 
\begin{align}
p_\mathrm{max} = 1-\left(\frac{3}{385}\right)^{\frac{4}{5}} \approx 0.9744.
\end{align}
We thus only accept the random number $p$ if it is smaller or equal than $p_\mathrm{max}$. This ensures that we obtain a velocity power law distribution with a slope of $\beta=1.8$ and a $14$ percent fraction of stars with velocity kicks of more than $30$ km s$^{-1}$.

\subsubsection{The Walkaway Case: Maxwellian velocity distribution}
The second scenario that we adopt for velocity kicks is designed to mimic the more conservative case of so called walkaway stars, where the bulk of the stars in the velocity distribution is moving at a speed lower than 10 km s$^{-1}$, following a velocity distribution that is consistent with a Maxwell-Boltzmann distribution:
\begin{align}
    p(v) \propto v^2 \exp{\left(-\frac{v^2}{2 \sigma^2}\right)}.
    \label{eq:maxwell}
\end{align}
where $\sigma$ is the width of the distribution. This is motivated based on previous work, both observational and theoretical, that shows some indication for a Maxwellian distribution \citep[e.g.][]{Eldridge2011, Silva2011}. However, we note that we take a rather conservative approach that is more consistent with previous research such as \citet{Renzo2019} which is suggesting that the bulk of runaway stars can actually be classified as walkaway stars. While such a distribution is a little harder to sample than the power law from the previous section, it becomes almost trivial when considering that the distribution of equation \ref{eq:maxwell} is the three dimensional representation of the Maxwell-Boltzmann distribution integrated over solid angle in spherical coordinates. Hence, this distribution in Cartesian coordinates consists of three (independent) Gaussian distributions. The latter can be straightforwardly sampled by a Gaussian random number generator, which is available in the \textit{gsl} package as the Ziggurat-algorithm, and is the fastest way of sampling the distribution of equation \ref{eq:maxwell}. We note that we explored different options to sample this distribution such as the Box-Muller method that can obtain Gaussian random numbers by utilising a linear combination of four uniform random numbers. In practice the difference is marginal, but the Ziggurat-algorithm is much faster. We note that for this model we also adopt a probability for a star to get any velocity kick. This is motivated by the fact a fraction of stars will be born as single stars and a fraction will be born as binary stars. In this model we assume that only binary stars can receive a velocity kick and we adopt a binary star fraction of 50 percent\footnote{The choice of 50 per cent is rather ad-hoc since most the exact fraction of runway/walkaway stars is not exact.}. We note that we did not assume this in the first model, to have a model close to the results of \citet{Oh2016} and the implementation of \citet{Andersson2020}.
Note that we do not explicitly sample the binary IMF, which is a clear limitation of this model. Since this model is supposed to represent the more conservative walkaway case we adopt a relatively low value of $\sigma =32$ km s$^{-1}$. 

\subsection{Initial conditions and simulations}

In our set of simulations we use the initial conditions of an isolated dwarf galaxy with a baryon mass of $2 \cdot 10^7$ M$_{\odot}$ and a dark matter mass of $2 \cdot 10^{10}$ M$_{\odot}$, which is an analogue of a classic Wolff-Lundmark-Moyette (WLM) system. The mass resolution is set to $4$ M$_{\odot}$. The force softening for this system is $0.1$ pc. While our choice for the gravitational softening is rather aggressive we note that extensive testing of different values of this parameter has little effect on global galactic properties such as morphology, star formation, and outflow rates. However, we note that a smaller choice of the softening-length than $0.1$ leads to sub-optimal gravity time stepping behaviour due to runaway star formation in dense regions that results in the gravitational ejection of single stars to arbitrarily large velocity. This effect is mainly significant in runs without PI-radiation (not presented here). The softening is roughly comparable to the size of the kernel that we observe in the dense regions of the ISM during the simulation. The initial conditions are generated with the code presented in \citet{springel05b} often referred to as ``MakeDiskGal'' or ``Makegalaxy''.\\ 
In total we carry out four simulations \textit{WLM-fiducial}, \textit{WLM-RunP} (power law runaway stars), \textit{WLM-RunM} (Maxwellian runaways), and \textit{WLM-inplane} (runaways only in xy plane) summarised in Table~\ref{tab:models}. The first case \textit{WLM-fiducial} represents a baseline case without runaway stars. The second case \textit{WLM-RunP} represents a strong runaway case. The third case \textit{WLM-RunM} represents a weaker runaway scenario, in which stars do not receive such high velocity kicks as in the \textit{WLM-RunP} scenario.  The final case \textit{WLM-inplane} samples the same power-law velocity distribution as \textit{WLM-RunP}, but kicks are only imparted in the x-y plane. This allows us to test the impact of runaways migrating within the plane of the galaxy versus escaping above or below the disk. 

\subsection{Definitions for outflow rates, inflow rates, loading factors and metal enrichment factors}
We follow the definitions for outflow rate, inflow rate, mass loading and enrichment factors of previous work, specifically, the definitions presented in \citet{Hu2019}. Generally, galactic wind quantities can be split into mass-flux (m), momentum-flux (p) and energy-flux (e) defined via:
\begin{align}
    {\bf \mathcal{F}}_\mathrm{m} = \rho \textbf{v},
\end{align}
\begin{align}
    {\bf \mathcal{F}}_\mathrm{p} = \rho \textbf{v} \cdot \textbf{v} + P,
\end{align}
\begin{align}
    {\bf \mathcal{F}}_\mathrm{e} = (\rho e_\mathrm{tot} + P)\textbf{v},
\end{align}
with $\rho$, $v$ and $P$ given as the fundamental fluid variables density, velocity and pressure. At all times $e_\mathrm{tot}$ is defined as the the total energy per unit mass $e_\mathrm{tot} = 0.5 \cdot \mathbf{v}^2 + u$, where $u$ is the internal energy per unit mass. In standard code units $u$ takes the units $km s^{-2}$. We note this to avoid confusion with converting these units to proper cgs units in the sections below. 

In this work, if not stated otherwise, all the flow rates are measured at a height of $1$ kpc above the midplane of the galaxies in a plane-parallel patch with thickness $dr$, where $dr=0.1$ kpc. 

We will adopt the following definitions for in and outflows rates, with the total flow rates defined via $\dot{M} = \dot{M}_\mathrm{out} - \dot{M}_\mathrm{in}$, $\dot{p} = \dot{p}_\mathrm{out} - \dot{p}_\mathrm{in}$ and $\dot{E} = \dot{E}_\mathrm{out} - \dot{E}_\mathrm{in}$. Outflow is defined by a positive value of the radial velocity given via $\mathbf{v \cdot \hat{n}} = v_{r} > 0$, and inflow is defined as the material in the patch with negative radial velocity $v_{r} < 0$. Hence, we can write down the discrete inflow and outflow rates for mass, momentum and energy:
\begin{align}
    \dot{M}_\mathrm{out} = \sum_{i, v_{i,r} > 0} \frac{m_{i}v_{i,r}}{dr},
\end{align}
\begin{align}
    \dot{M}_\mathrm{in} = - \sum_{i, v_{i,r} < 0} \frac{m_{i}v_{i,r}}{dr},
\end{align}
\begin{align}
    \dot{p}_\mathrm{out} = \sum_{i, v_{i,r} > 0} \frac{m_{i} [v_{i,r}^2 + (\gamma -1) u_{i}]}{dr},
\end{align}
\begin{align}
    \dot{p}_\mathrm{in} = - \sum_{i, v_{i,r} < 0} \frac{m_{i} [v_{i,r}^2 + (\gamma -1) u_{i}]}{dr},
\end{align}
\begin{align}
    \dot{E}_\mathrm{out} = \sum_{i, v_{i,r} > 0} \frac{m_{i} [v_{i,r}^2 + \gamma u_{i}]v_{i,r}}{dr},
\end{align}
\begin{align}
    \dot{E}_\mathrm{in} = - \sum_{i, v_{i,r} < 0} \frac{m_{i} [v_{i,r}^2 + \gamma u_{i}]v_{i,r}}{dr},
\end{align}
where we use $P = \rho u (\gamma -1)$ as an equation of state. Furthermore, we will compute metal outflow rates via:
\begin{align}
    \dot{M}_\mathrm{out} = \sum_{i, v_{i,r} > 0} \frac{Z_{i} m_{i}v_{i,r}}{dr},
\end{align}
\begin{align}
    \dot{M}_\mathrm{in} = - \sum_{i, v_{i,r} < 0} \frac{Z_{i} m_{i}v_{i,r}}{dr},
\end{align}
where $Z_{i}$ is the metallicity of particle $i$.
It is useful to investigate these parameters normalised to a set of reference quantities, defined as the so-called loading factors for mass ($\eta_\mathrm{m}$), metals ($\eta_\mathrm{Z}$), momentum ($\eta_\mathrm{p}$) and energy ($\eta_\mathrm{e}$) and given by:
\begin{enumerate}
    \item outflow mass loading factor: $\eta_\mathrm{m}^\mathrm{out} = \dot{M}_\mathrm{out} / \overline{\mathrm{SFR}}$,
    \item inflow mass loading factor: $\eta_\mathrm{m}^\mathrm{in} = \dot{M}_\mathrm{in} / \overline{\mathrm{SFR}}$,
    \item momentum outflow loading factor: $\eta_\mathrm{p}^\mathrm{out} = \dot{p}_\mathrm{out}/(p_\mathrm{SN} R_\mathrm{SN})$,
    \item momentum inflow loading factor: $\eta_\mathrm{p}^\mathrm{in} = \dot{p}_\mathrm{in}/(p_\mathrm{SN} R_\mathrm{SN})$,
    \item energy outflow loading factor: $\eta_\mathrm{e}^\mathrm{out} = \dot{E}_\mathrm{out}/(E_\mathrm{SN} R_\mathrm{SN})$,
    \item energy inflow loading factor: $\eta_\mathrm{e}^\mathrm{in} = \dot{E}_\mathrm{in}/(E_\mathrm{SN} R_\mathrm{SN})$,
    \item outflow metal loading factor: $\eta_\mathrm{Z}^\mathrm{out} = \dot{M}_\mathrm{Z, out} / (m_\mathrm{Z} R_\mathrm{SN})$,
    \item inflow metal loading factor: $\eta_\mathrm{Z}^\mathrm{in} = \dot{M}_\mathrm{Z, in} / (m_\mathrm{Z} R_\mathrm{SN})$,    
\end{enumerate}
where we adopt $E_\mathrm{SN} = 10^{51}$ erg, p$_{SN} = 3 \cdot 10^{4}$ M$_{\odot}$ km s$^{-1}$, the supernova rate R$_{SN} = \overline{\mathrm{SFR}}/(100 M_{\odot})$ and the metal mass (IMF-averaged) $m_\mathrm{Z} = 2.5 M_{\odot}$ which we take for best comparison with previous dwarf galaxy studies from \citet[][]{Hu2019}.
We compute the mean star formation rate $\overline{\mathrm{SFR}}$ over the time span of one Gyr, which is the total run time of the simulation. 
Additionally, we define metal-enrichment factors that not only quantify the amount of metals transported in the outflows, but can quantify to what degree metals are over- or under-abundant in galaxy outflows relative to the ISM. This can be achieved by normalising the metal outflow rate to the mass outflow rate, weighted by the background metallicity of the galactic ISM, Z$_\mathrm{gal} = 0.1 Z_{\odot}$: 
\begin{enumerate}
    \item outflow enrichment factor: $y_\mathrm{Z}^\mathrm{out} = \dot{M}_\mathrm{Z, out}/ (\dot{M}_\mathrm{out} Z_\mathrm{gal})$, 
    \item inflow enrichment factor: $y_\mathrm{Z}^\mathrm{in} = \dot{M}_\mathrm{Z, out}/ (\dot{M}_\mathrm{in} Z_\mathrm{gal})$.
\end{enumerate}

\section{Results}
\label{sec:results}

In this section we will study the outflow properties of all of our dwarf galaxy simulations with a specific focus on the time evolution of the outflow rates and wind loading factors.

\begin{figure*}
    \centering
    \includegraphics[scale=0.4]{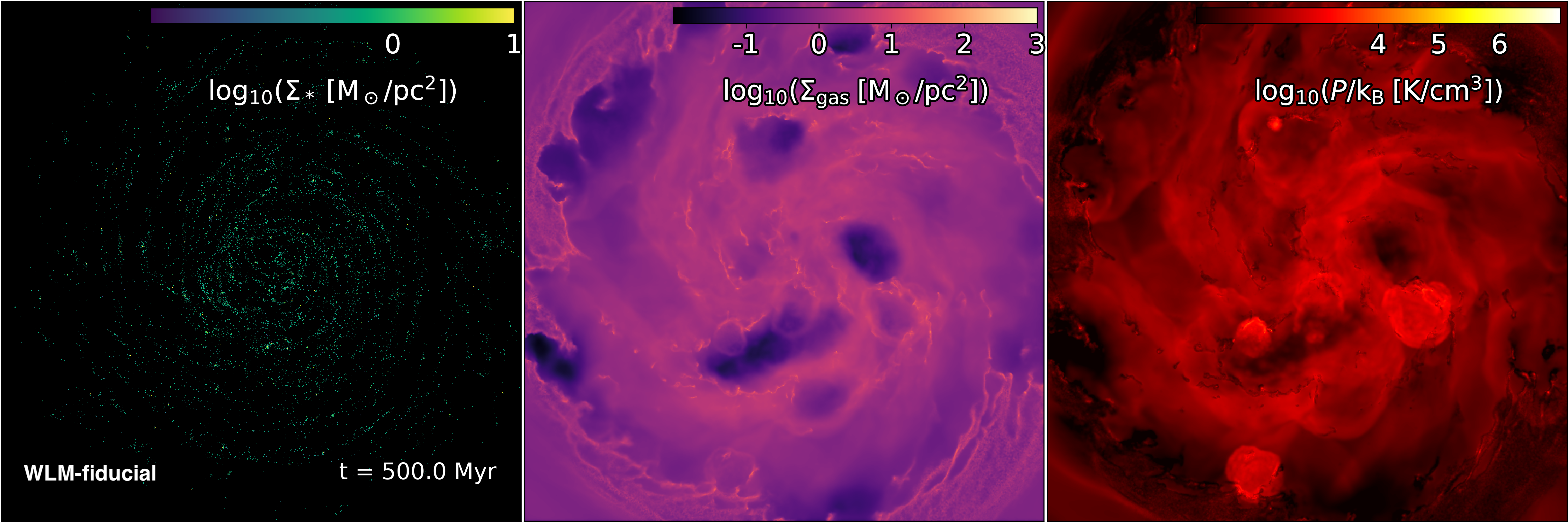}
    \includegraphics[scale=0.4]{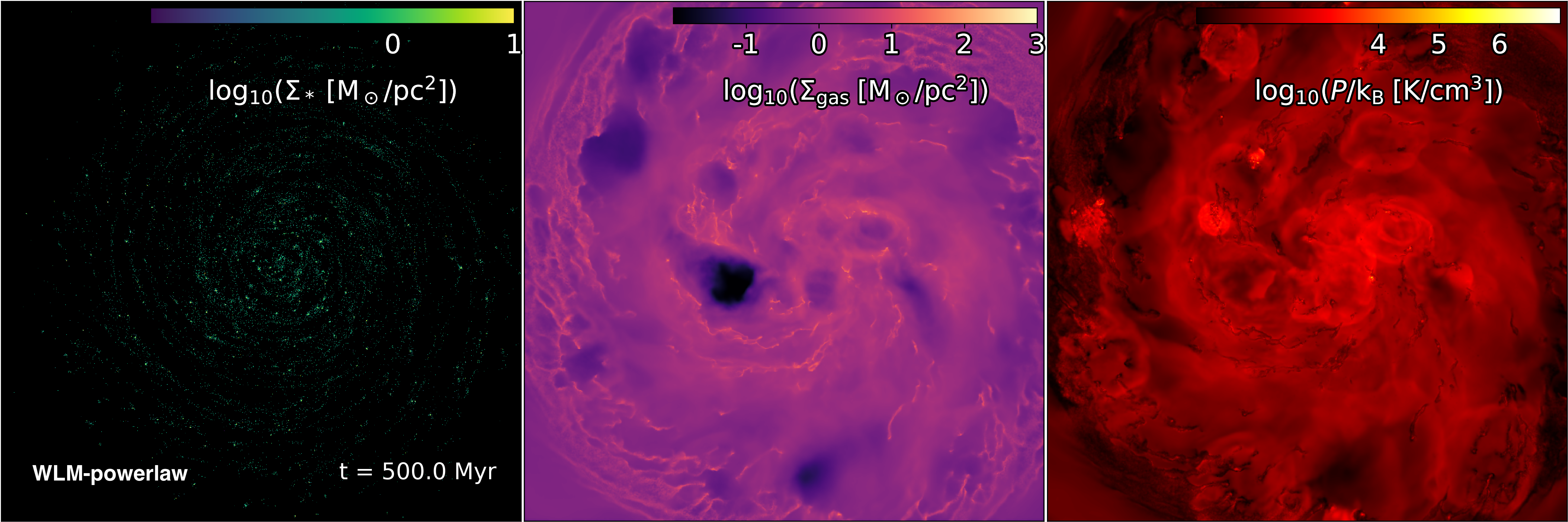}
    \includegraphics[scale=0.4]{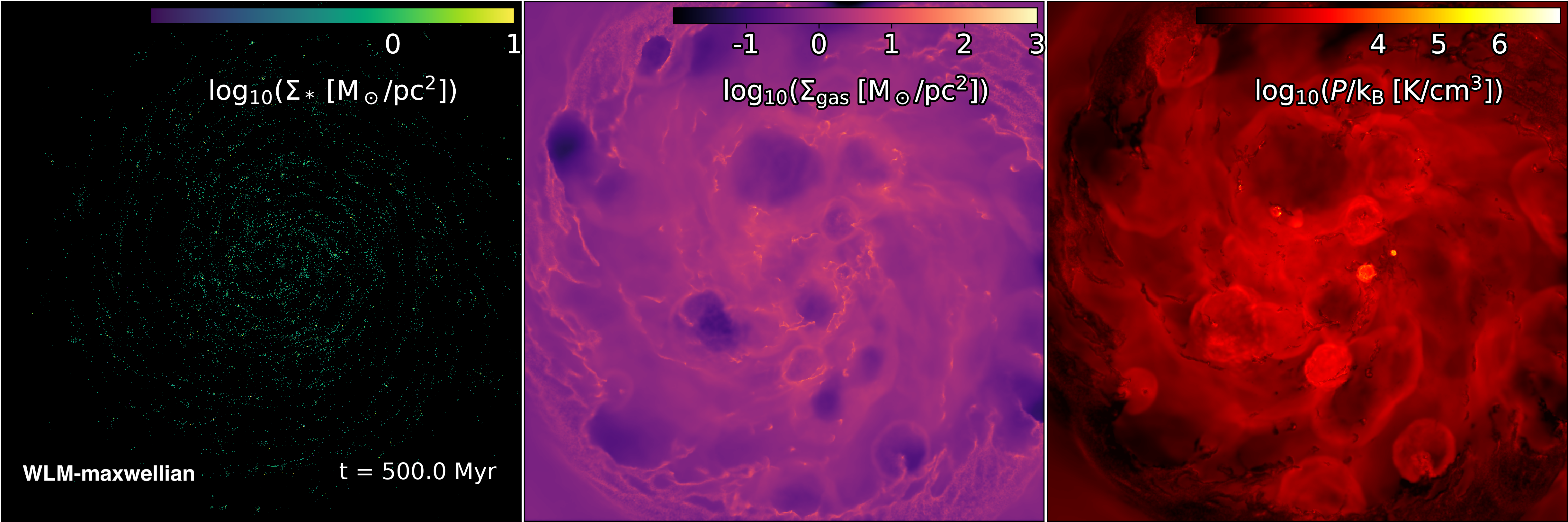}
    \includegraphics[scale=0.4]{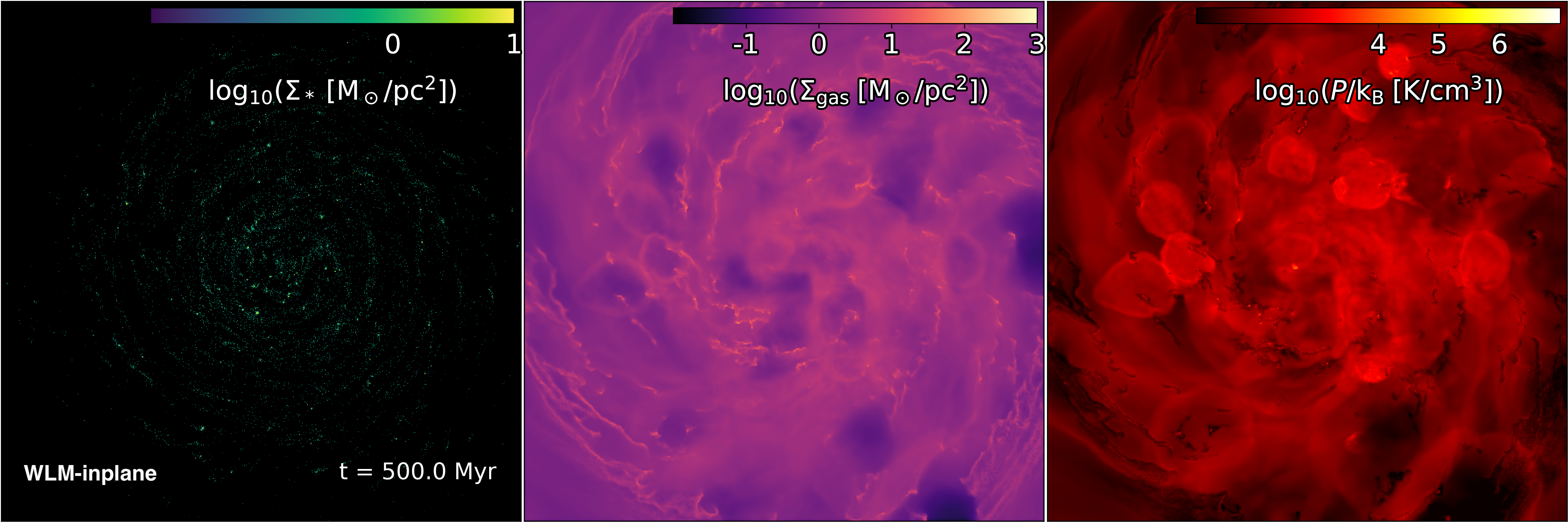}
    \caption{Visualisation of the the face-on stellar surface density (left), the gas surface density (center) and the pressure (right), for our isolated dwarf galaxy simulations, for the runs \textit{WLM-fiducial} (first row), \textit{WLM-RunP} (second row), \textit{WLM-RunM} (third row) and \textit{WLM-inplane} (fourth row). We find only very minor differences between the different runs in terms of the morphology of the stellar structure and the ISM.}
    \label{fig:faceon}
\end{figure*}

\begin{figure*}
    \centering
    \includegraphics[scale=0.4]{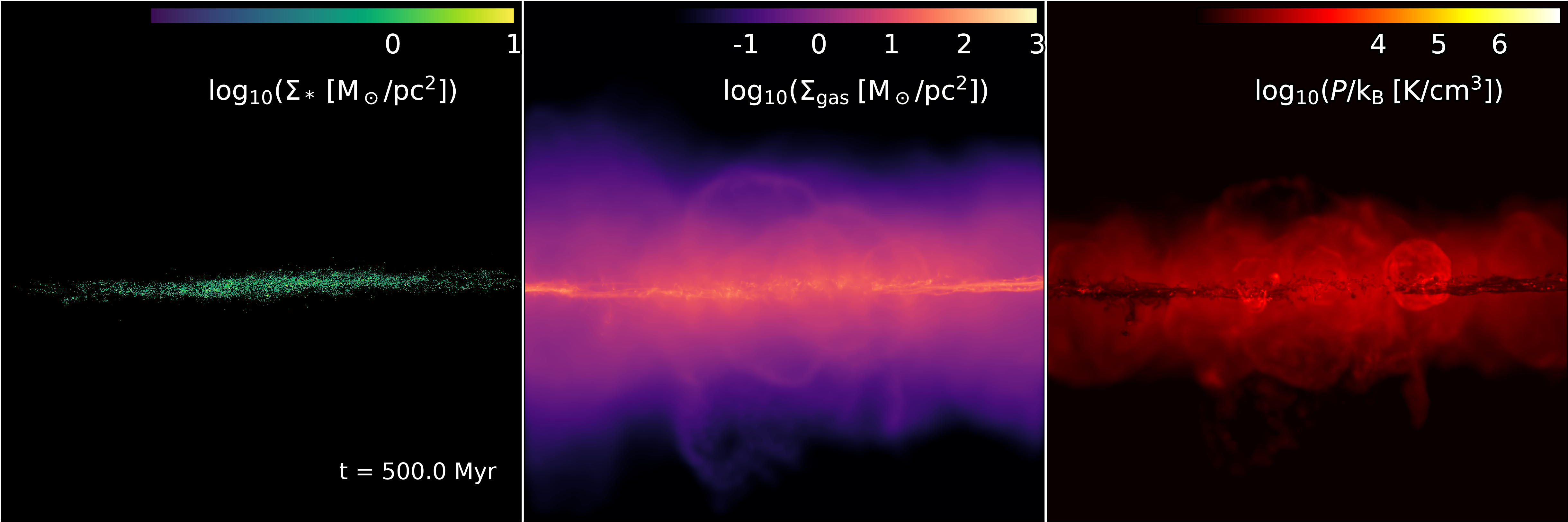}
    \includegraphics[scale=0.4]{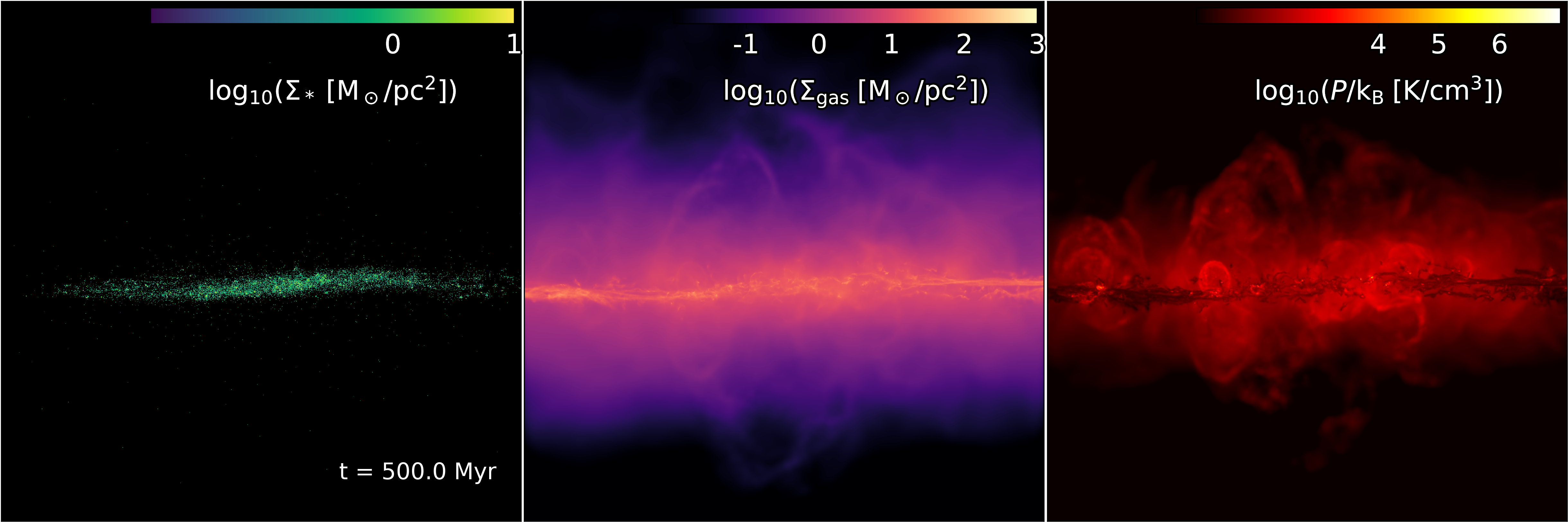}
    \includegraphics[scale=0.4]{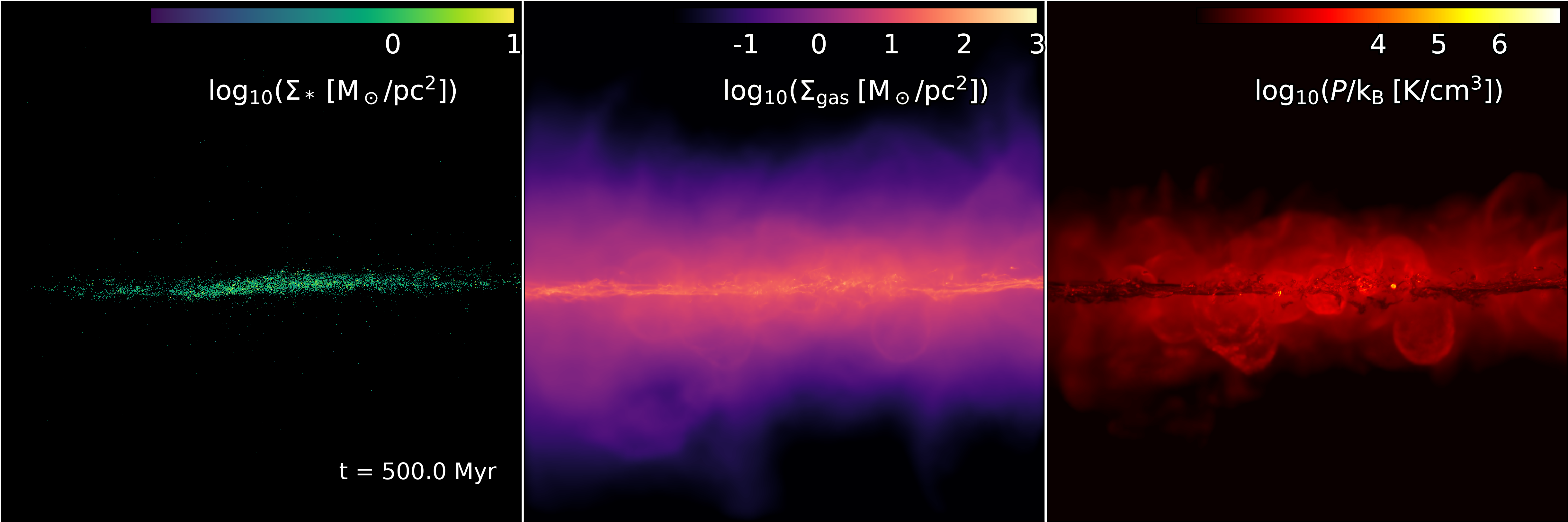}
    \includegraphics[scale=0.4]{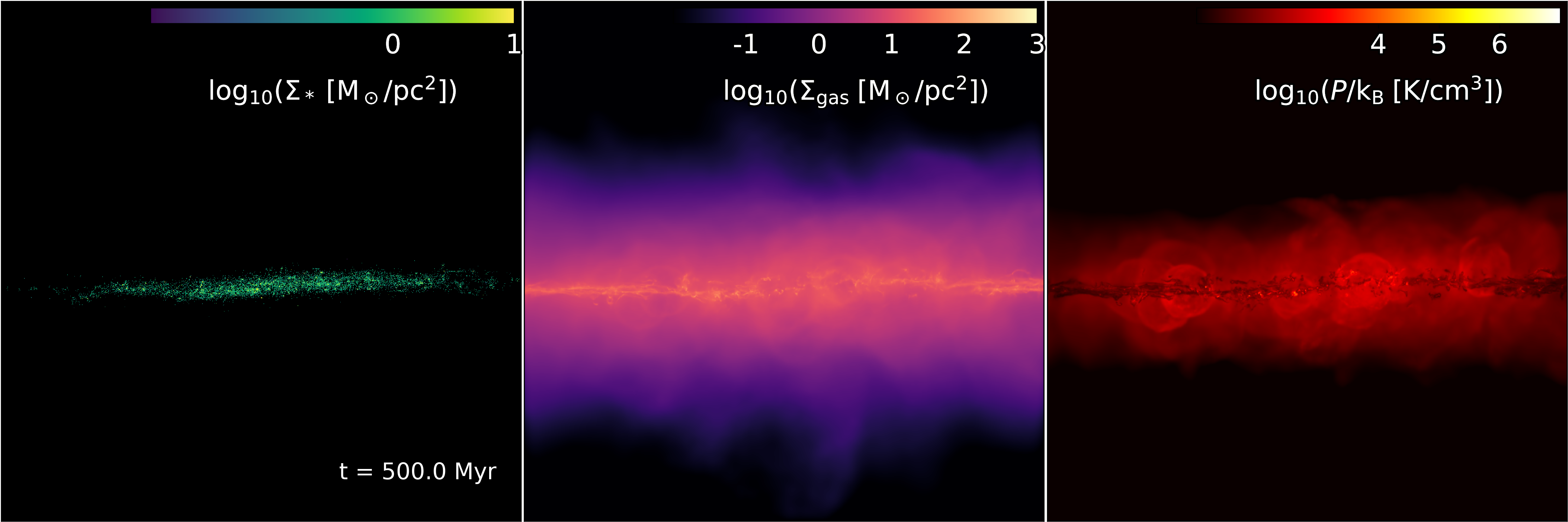}
    \caption{We visualise the edge-on stellar surface density (left), the gas surface density (center) and the temperature (right), for our isolated dwarf galaxy simulations, for the runs \textit{WLM-fiducial} (first row), \textit{WLM-RunP} (second row), \textit{WLM-RunM} (third row) and \textit{WLM-inplane} (fourth row). We find only very minor differences between the different runs in terms of the morphology of the stellar structure and the ISM. However, the stellar disc seems to be slightly ``puffed up'' in the runs \textit{WLM-RunP} and \textit{WLM-RunM} compared to the runs \textit{WLM-fiducial} and \textit{WLM-inplane}.}
    \label{fig:edgeon}
\end{figure*}

\begin{figure}
    \centering
    \includegraphics[scale=0.5]{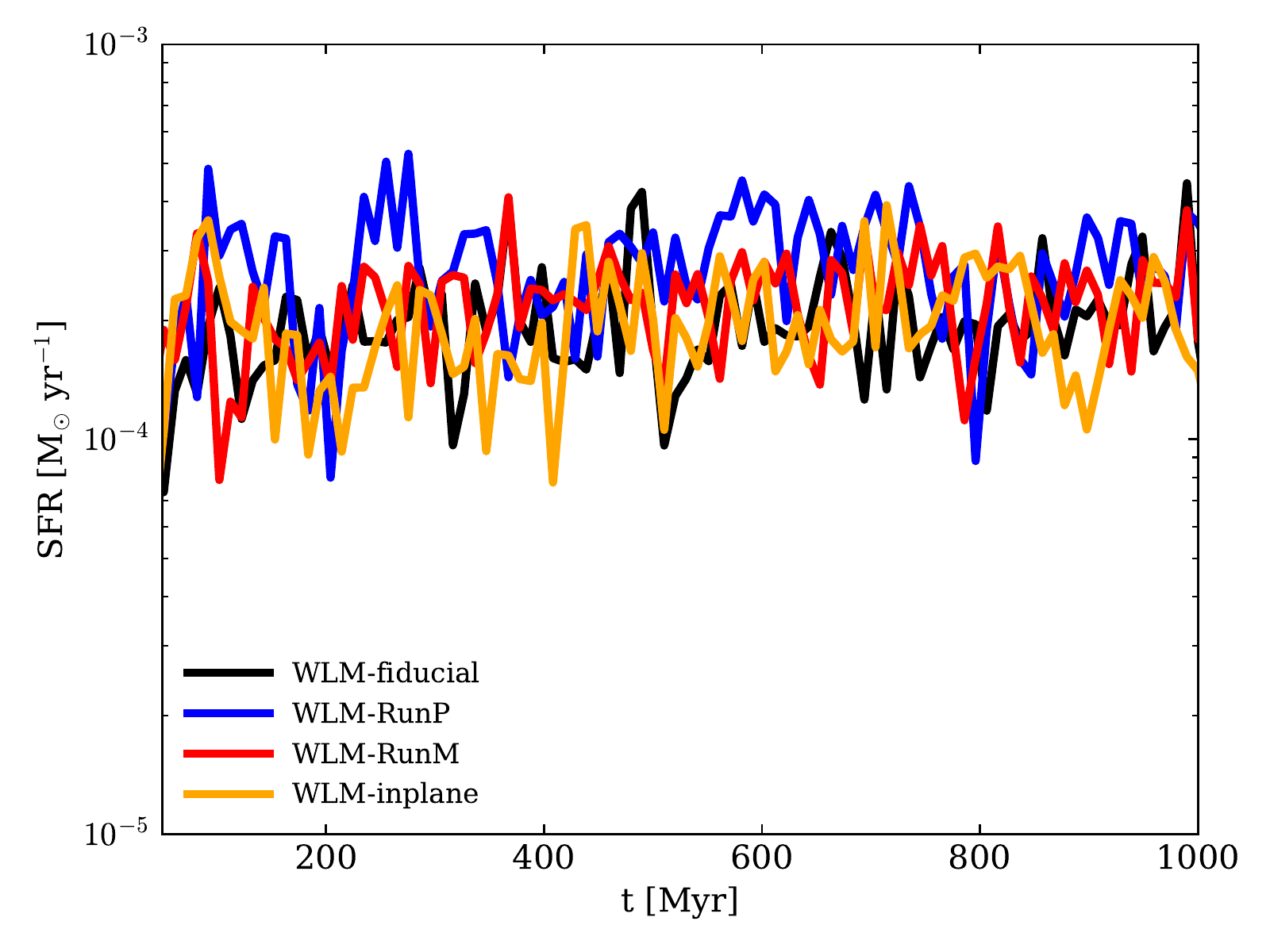}
        \includegraphics[scale=0.5]{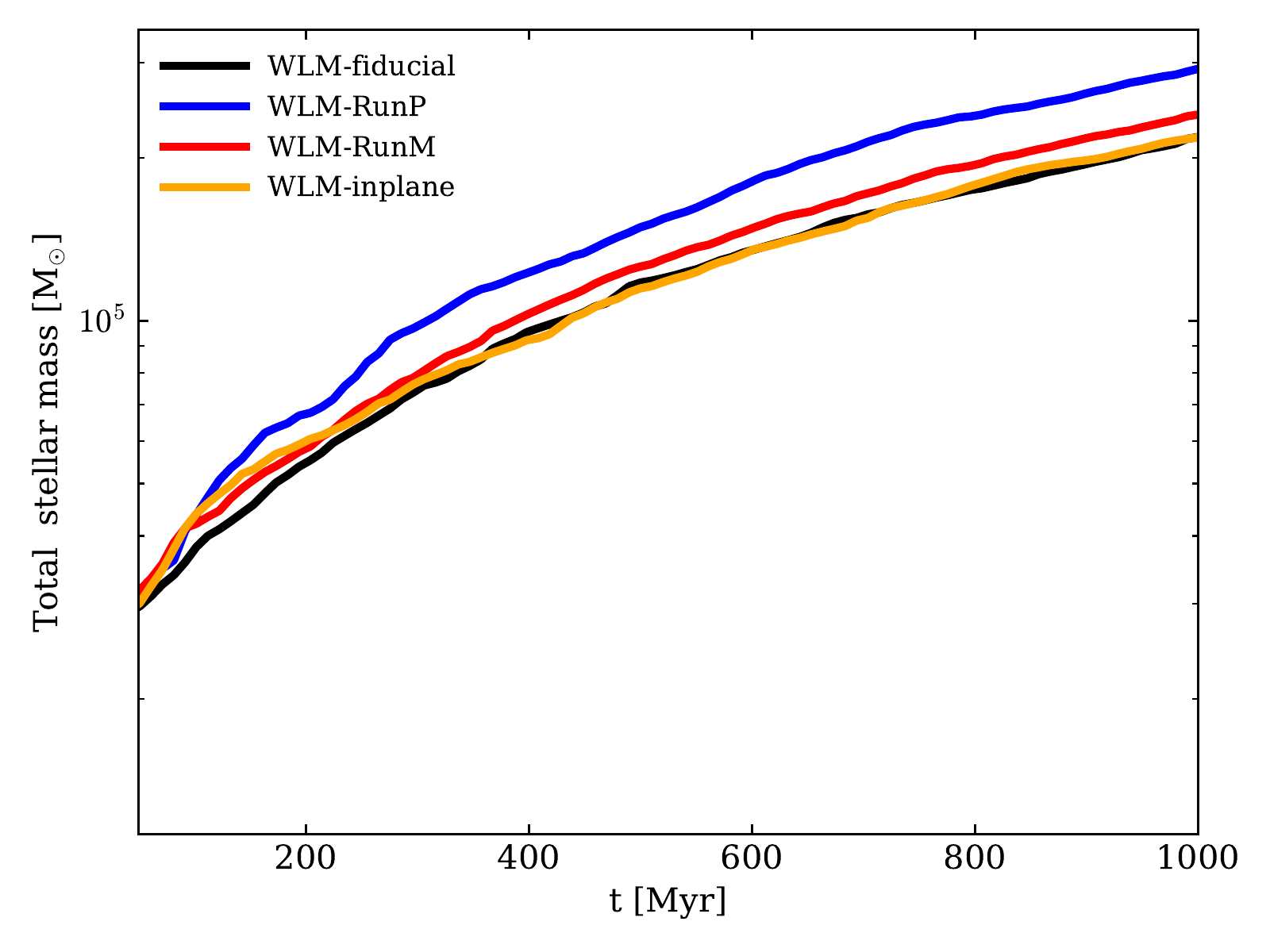}
    \caption{We show the star formation rate for the runs \textit{WLM-fiducial} (black), \textit{WLM-RunP} (blue), \textit{WLM-RunM} (red) and \textit{WLM-inplane} (orange) in the top panel, as well as the total stellar mass of the systems in the bottom panel as a function of time. We find a slight boost of the star formation rate in the runaway star models. This may be due to the reduction of feedback from PI-radiation as some massive stars quickly leave their natal clouds. However, the time averaged star formation rate is only weakly influenced by this effect due to the self-regulating nature of the star formation and feedback cycle.}
    \label{fig:sfr}
\end{figure}

\begin{figure}
    \centering
    \includegraphics[scale=0.5]{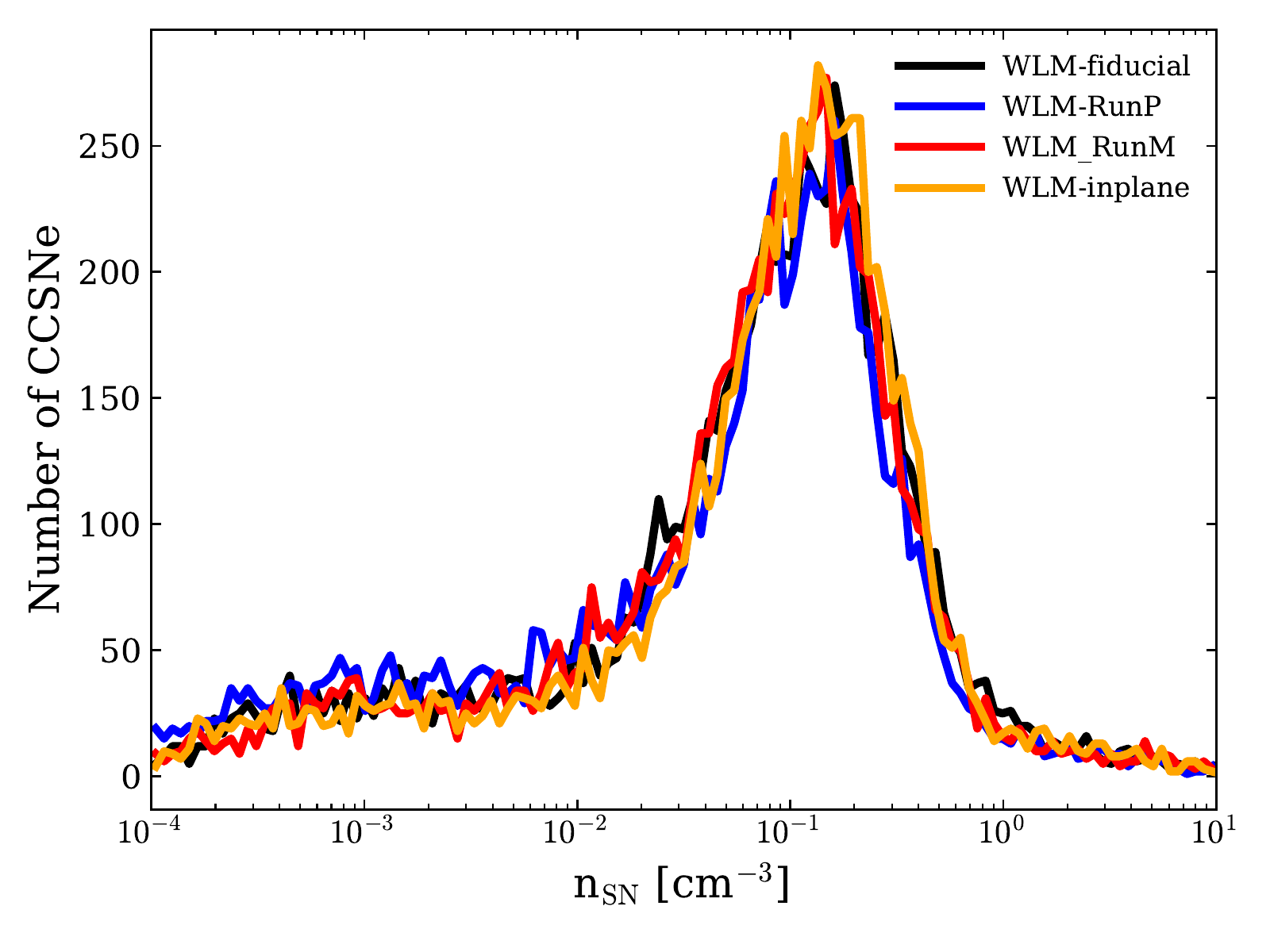}
    \caption{Distribution of the density of the gas in which SNe explode, for all runs \textit{WLM-fiducial} (black), \textit{WLM-RunP} (blue), \textit{WLM-RunM} (red) and \textit{WLM-inplane} (orange). For all runs we took a statistically large enough sample of 3000 SNe explosions to allow for direct comparison. The data are binned in 128 equally spaced logarithmic bins in the range from $10^{-4}$ to $10^{1}$ cm$^{-3}$. Runaway stars have only a weak effect on the distribution function but in the run \textit{WLM-RunP} there is a slight excess of stars exploding in lower density environments. However, this is within the model scatter, which implies that major changes in global properties are driven by the altitude above or below the disc at which the SNe explode and not the redistribution of the SNe in the galactic plane.}
    \label{fig:env_dens_runaways}
\end{figure}

\begin{figure}
    \centering
    \includegraphics[scale=0.5]{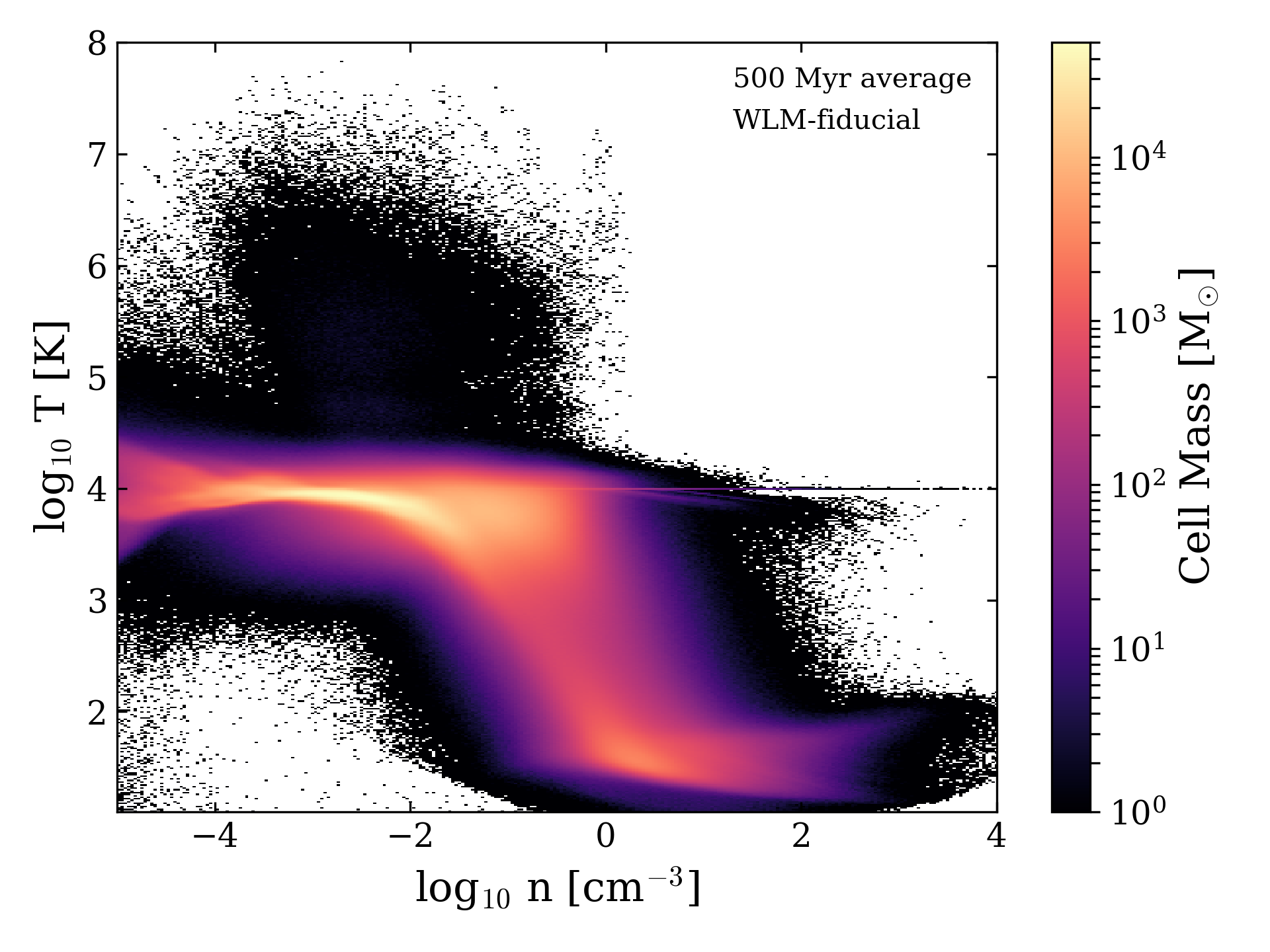}
    \includegraphics[scale=0.5]{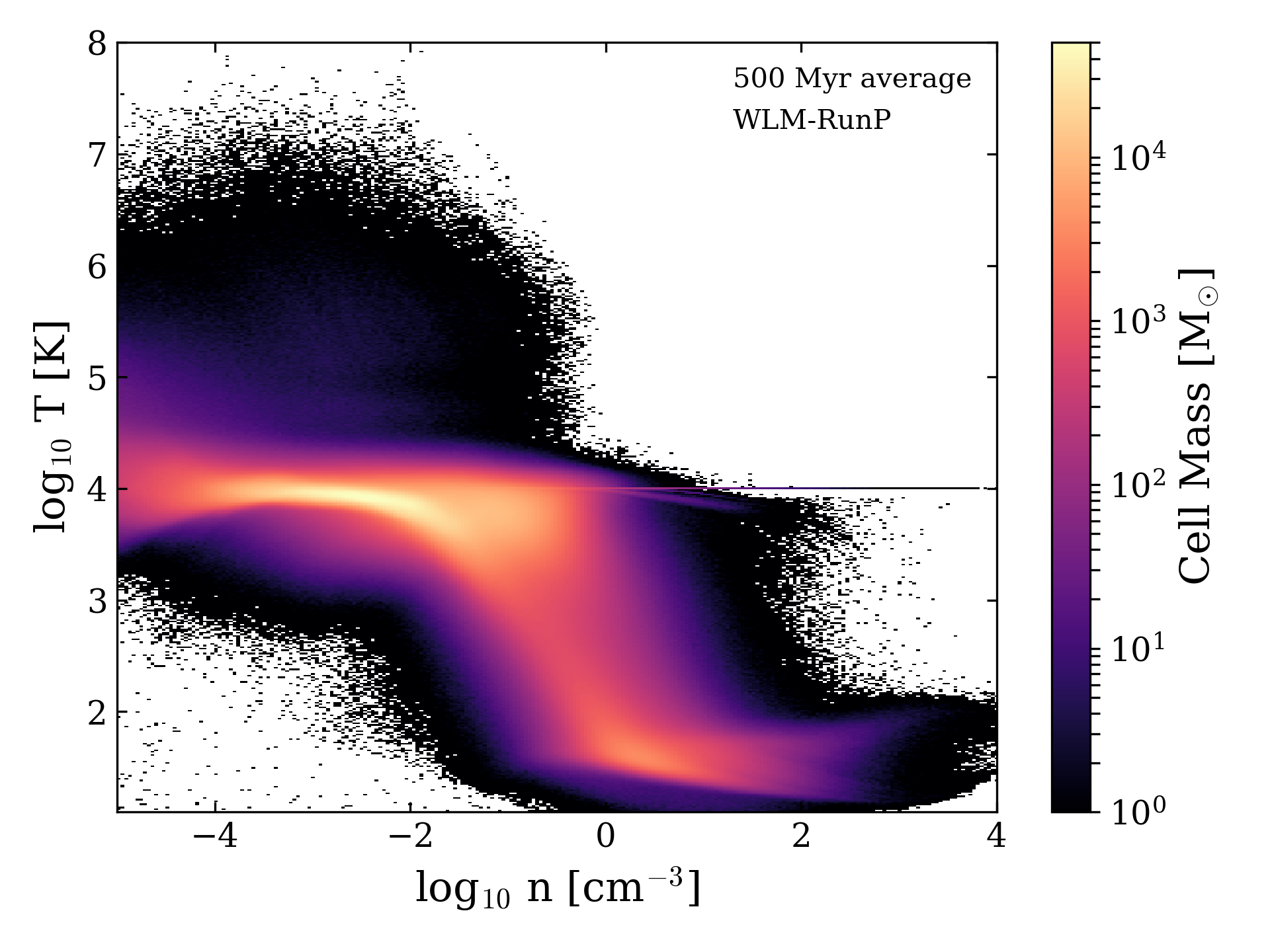}
    \includegraphics[scale=0.5]{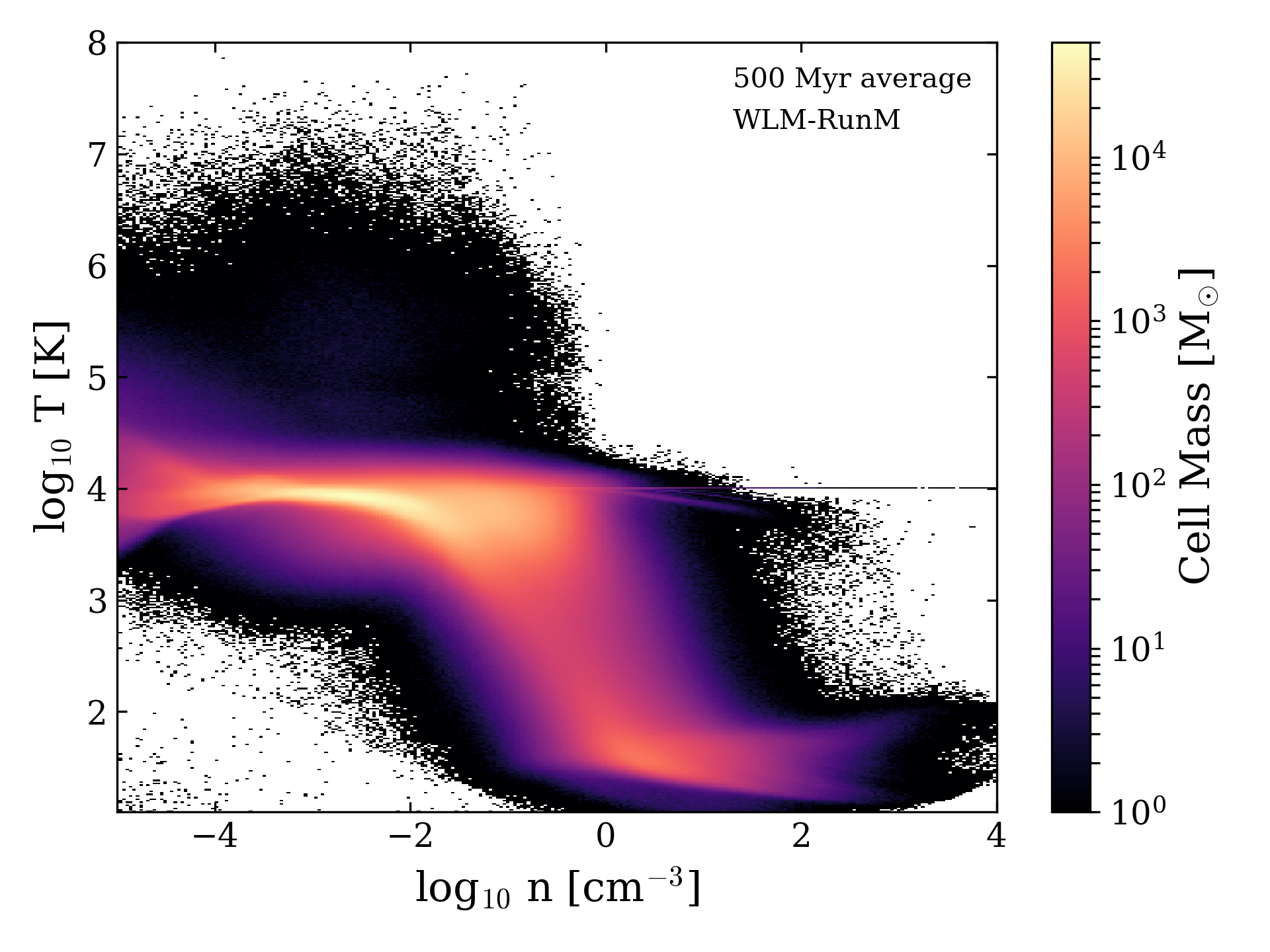}
    \caption{We show the time averaged mass weighted density-temperature phase space for the runs \textit{WLM-fiducial} (top), \textit{WLM-RunP} (centre) and \textit{WLM-RunM} (bottom). All plots are averaged over a time scale of 500 Myr. The runs \textit{WLM-RunP} and \textit{WLM-RunM} show excess mass in the hot phase of the ISM. However, since the hot phase typically does not contain most of the mass, the trend is more apparent in the volume weighted version of this figure (Fig.~\ref{fig:phase_diagrams_avg_vol}).
    }
    \label{fig:phase_diagrams_avg}
\end{figure}

\begin{figure}
    \centering
    \includegraphics[scale=0.5]{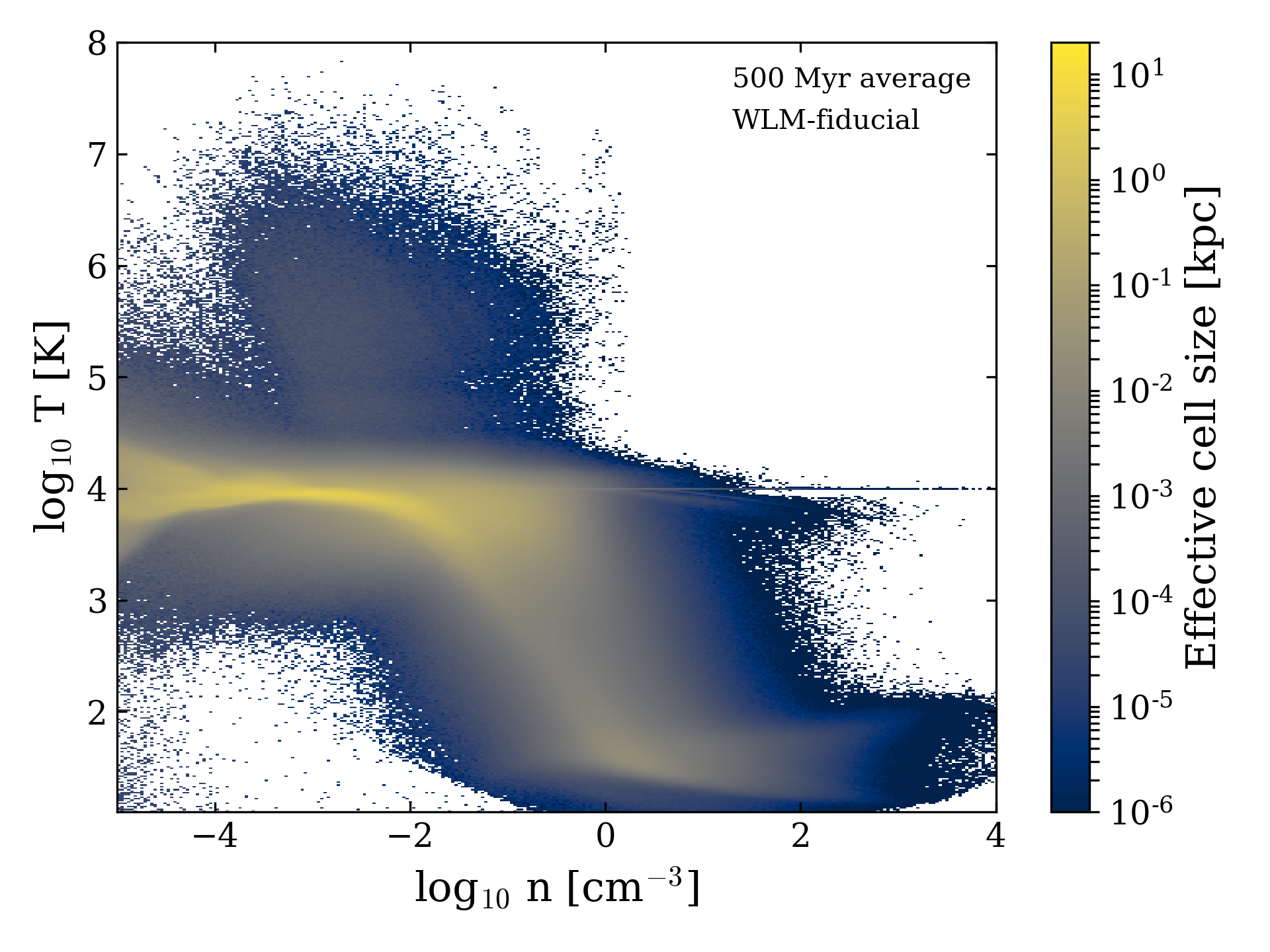}
    \includegraphics[scale=0.5]{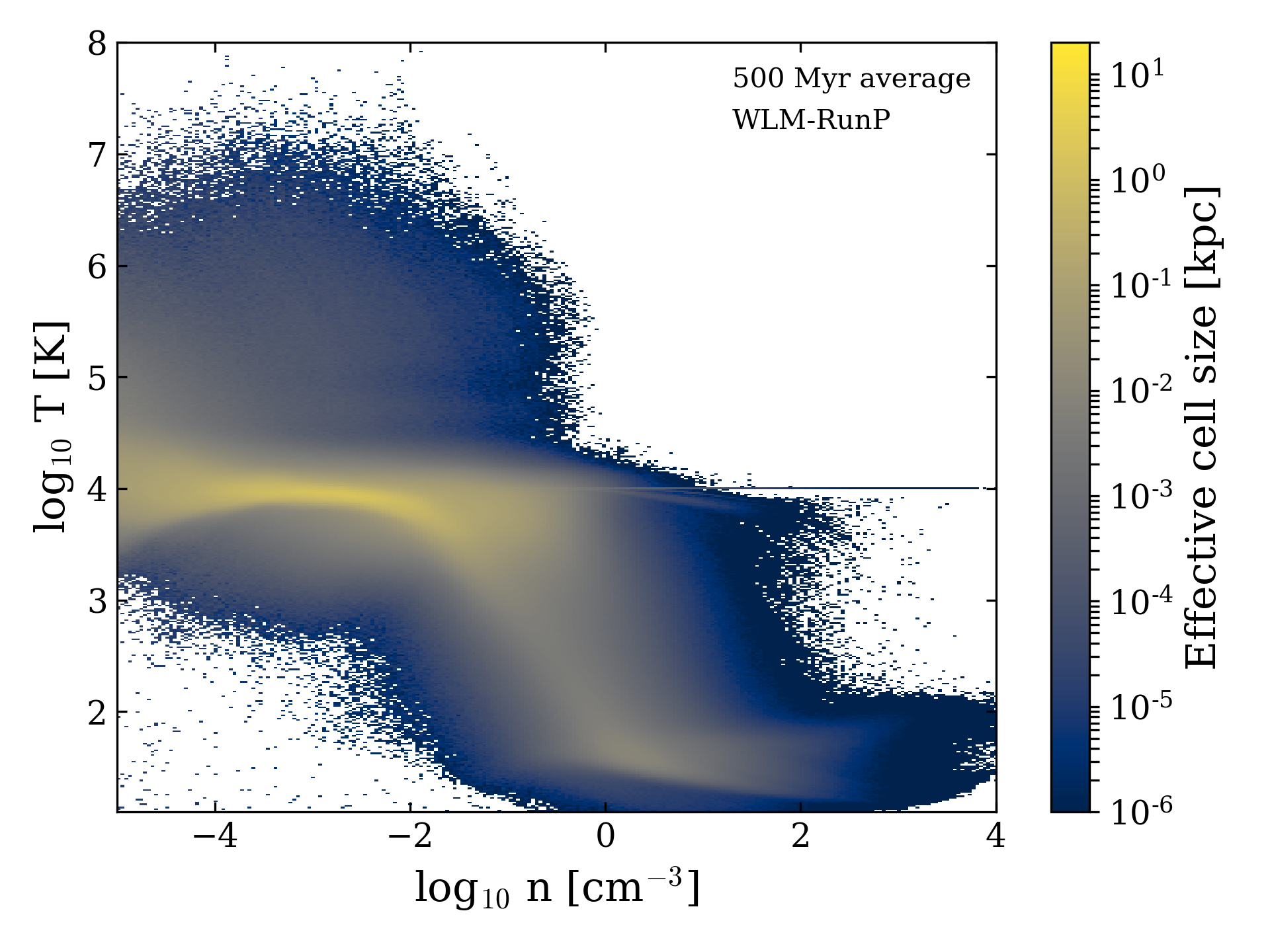}
    \includegraphics[scale=0.5]{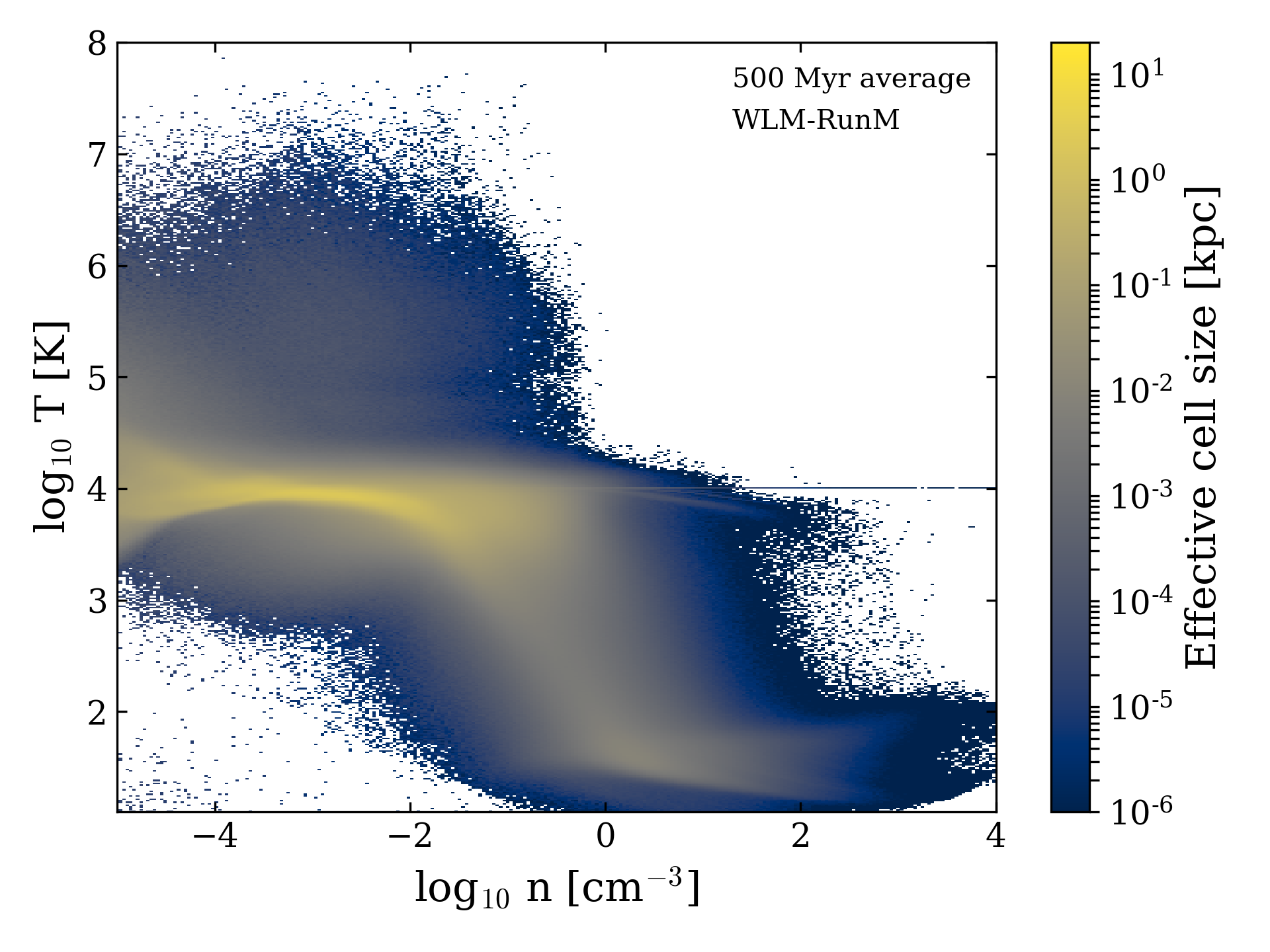}
    \caption{We show the time averaged volume weighted density-temperature phase space for the runs \textit{WLM-fiducial} (top), \textit{WLM-RunP} (centre) and \textit{WLM-RunM} (bottom). All plots are averaged over a time scale of 500 Myr. It is apparent that the inclusion of runaway stars can have a significant effect on establishing the hot, volume filling phase of the ISM.}
    \label{fig:phase_diagrams_avg_vol}
\end{figure}

\begin{figure}
    \centering
    \includegraphics[scale=0.5]{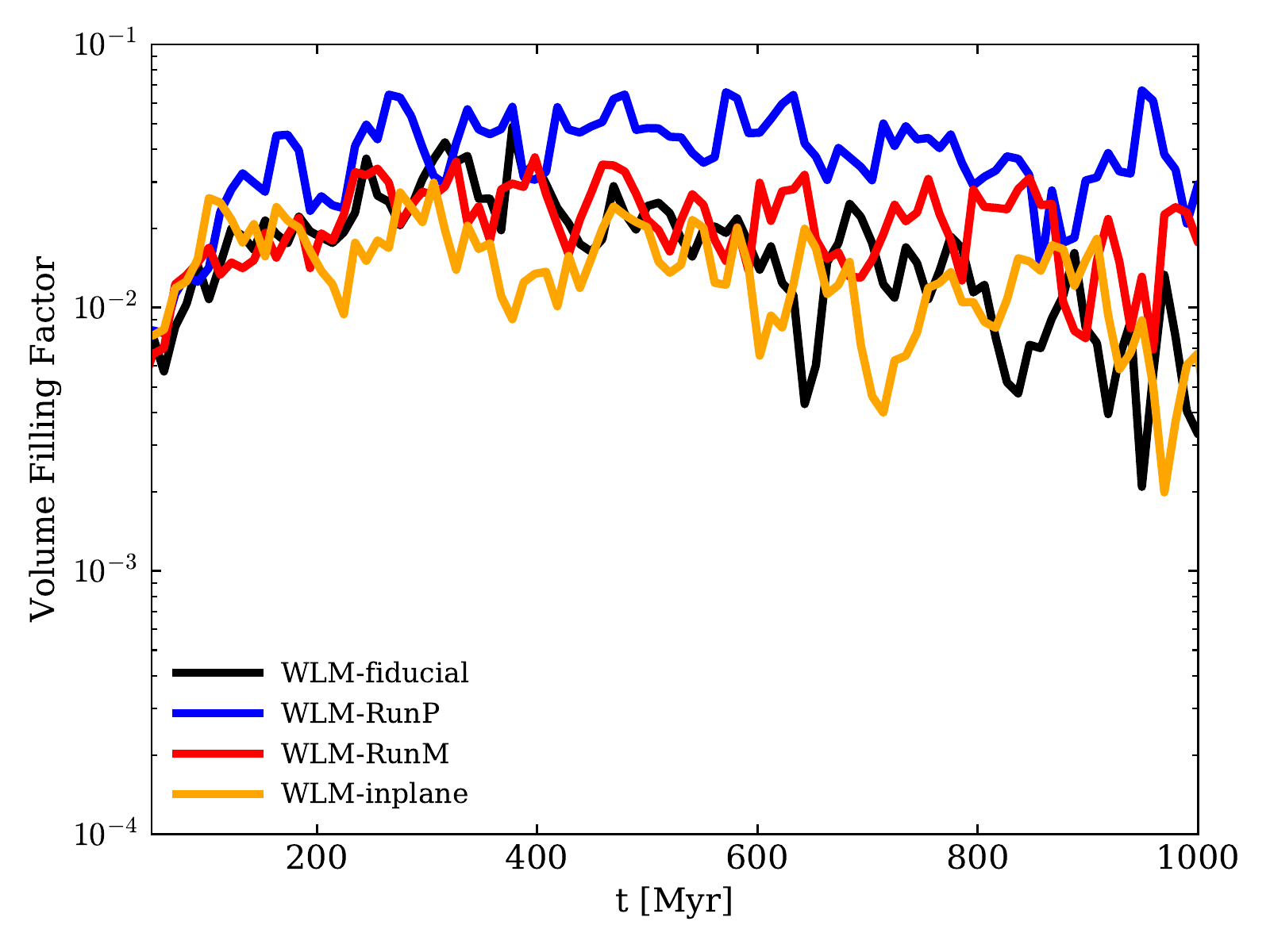}
    \caption{We show the volume filling factor of the hot gas (above a temperature of $3 \times 10^4$ K) as a function of time for all four models \textit{WLM-fiducial} (black), \textit{WLM-RunP} (blue), \textit{WLM-RunM} (red), and \textit{WLM-inplane} (orange). One can clearly see that the runaway models show an increase in the volume filling factor of the hot phase. As pointed out in previous studies of \citet{Hu2017} and \citet{Steinwandel2020}, the volume filling factor is critical for driving outflows in galaxies.}
    \label{fig:vff}
\end{figure}

\begin{figure}
    \centering
    \includegraphics[scale=0.5]{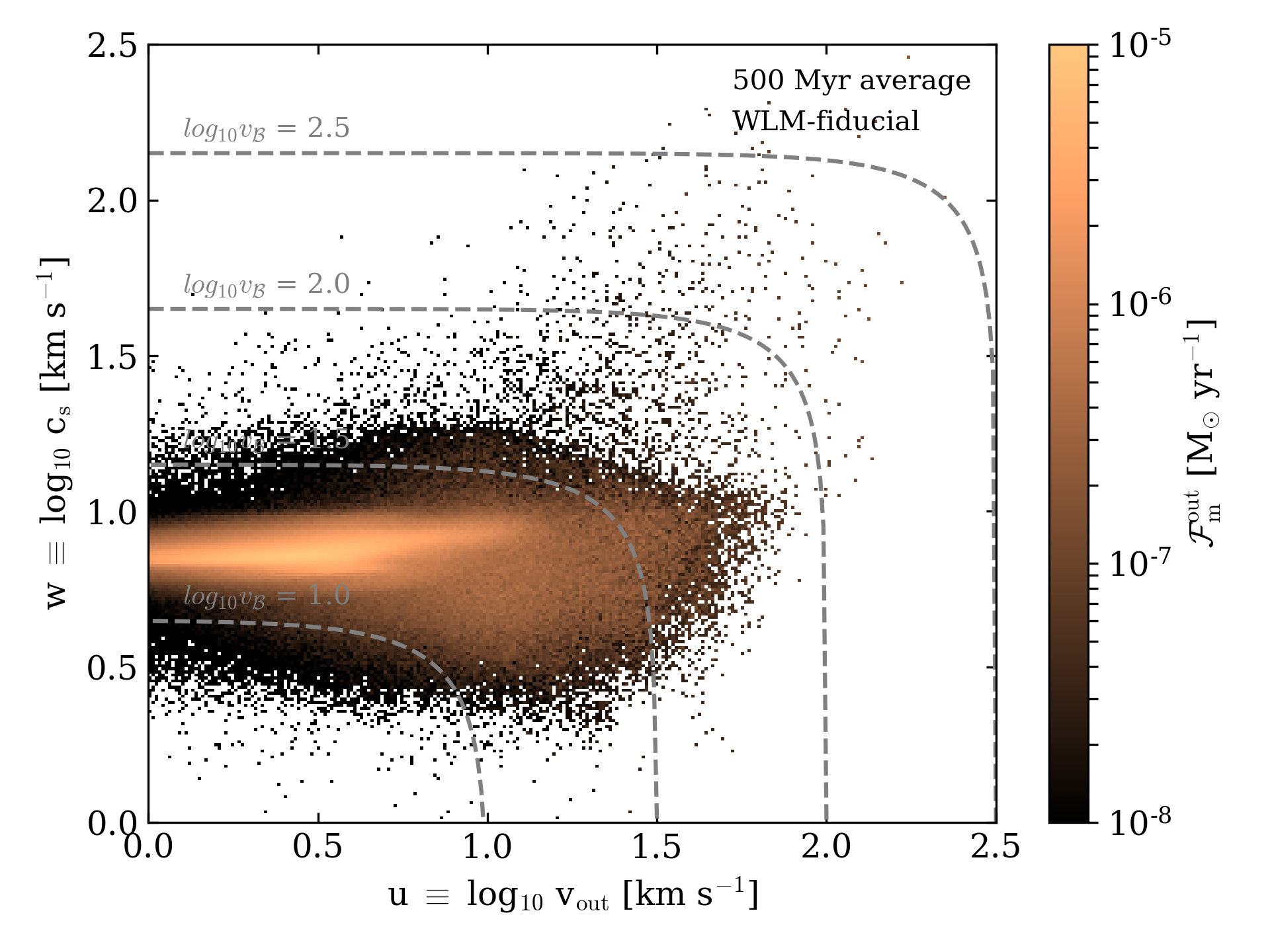}
    \includegraphics[scale=0.5]{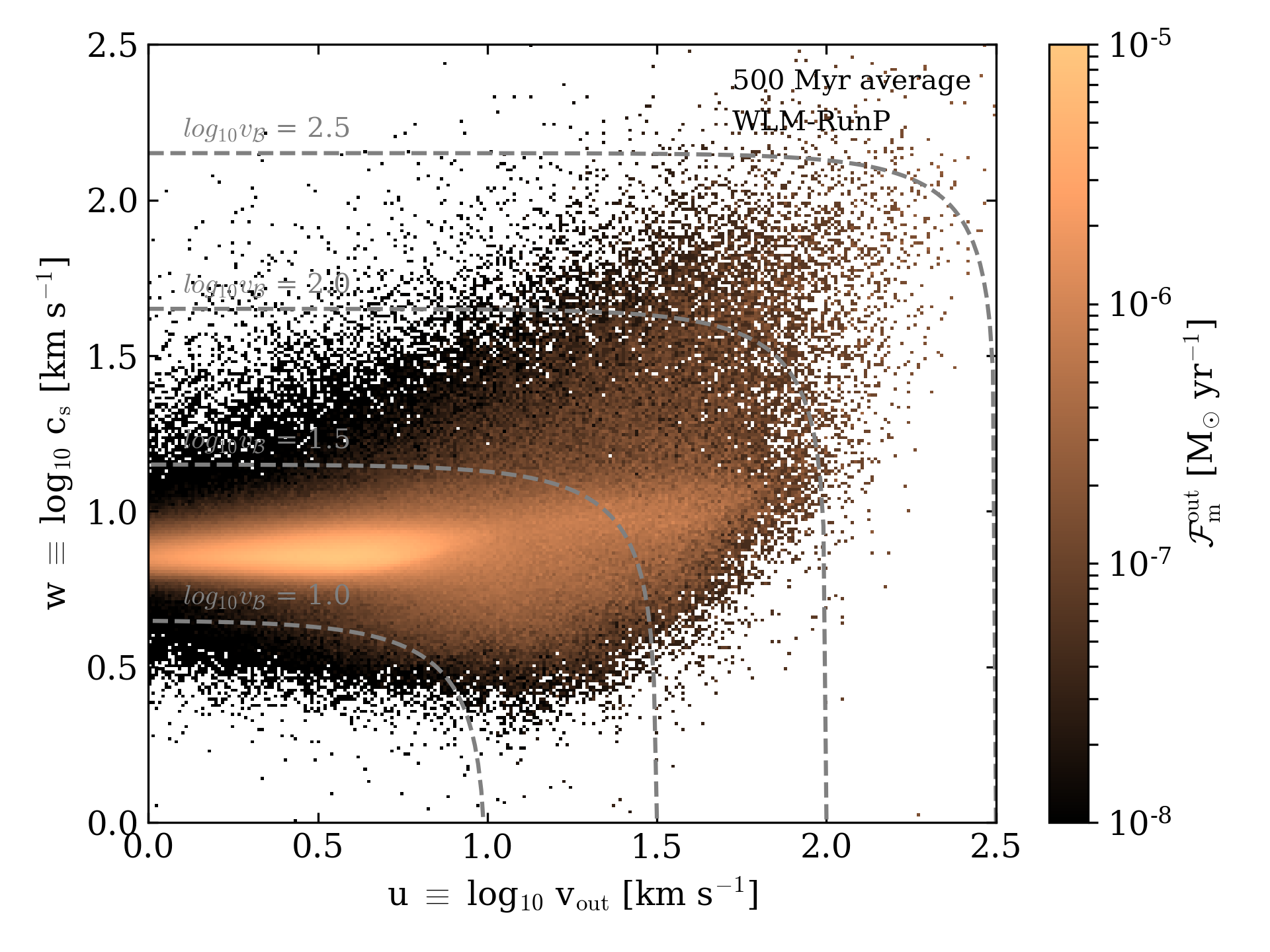}
    \caption{We show the outflow velocity as a function of the sound speed in logarithmic space for the models \textit{WLM-fiducial} (top) and \textit{WLM-RunP}. The colour code indicates the averaged outflow mass flux rate over a time span of 500 Myr. While these indicate that the bulk of the mass in the outflow is transported by the cold phase, we find a significant increase in mass transport in the hot phase of the wind, that extends to higher outflow velocities and sound speed for the simulation \textit{WLM-RunP} (bottom).}
    \label{fig:logu_mass}
\end{figure}

\begin{figure}
    \centering
    \includegraphics[scale=0.5]{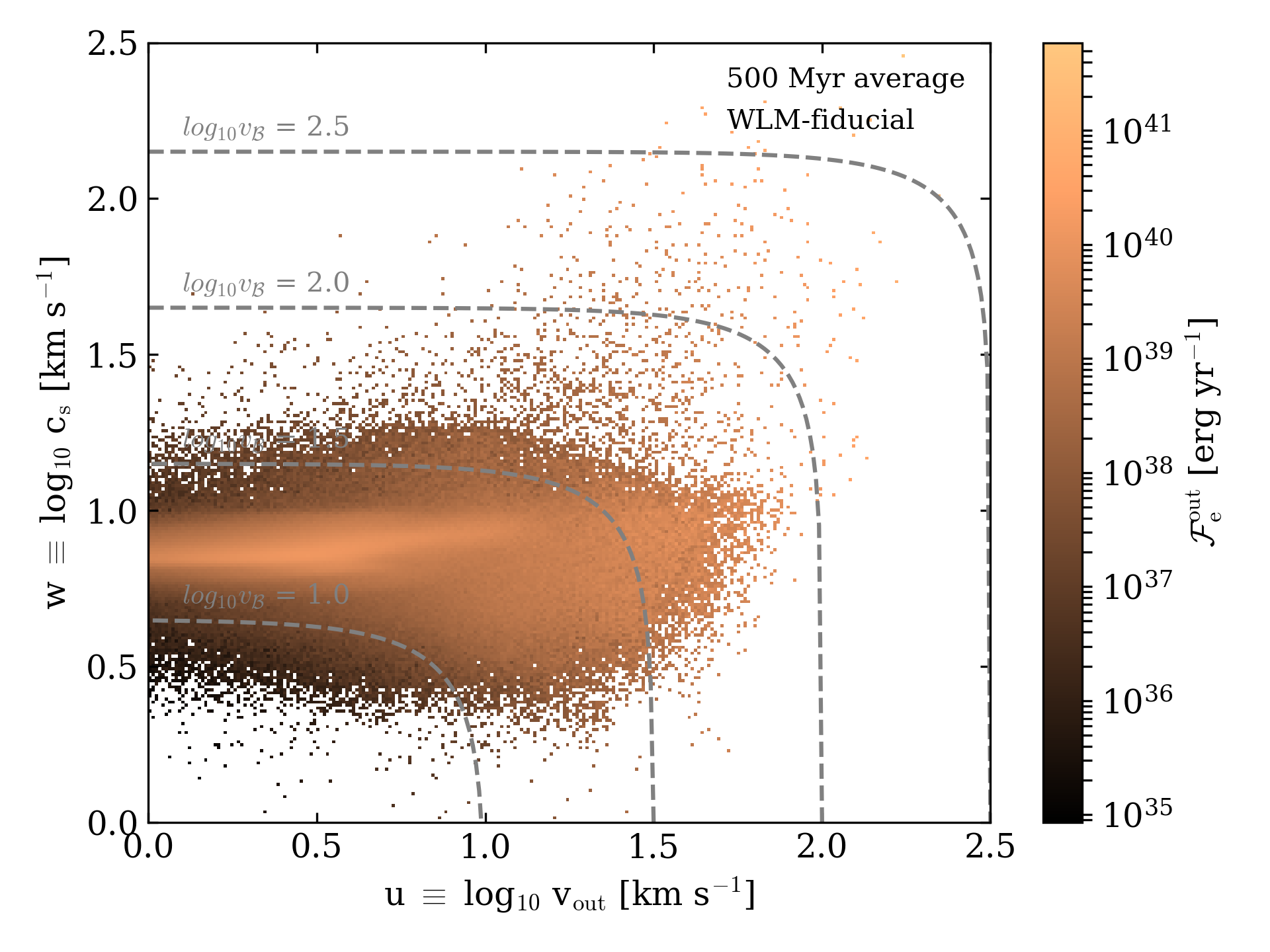}
    \includegraphics[scale=0.5]{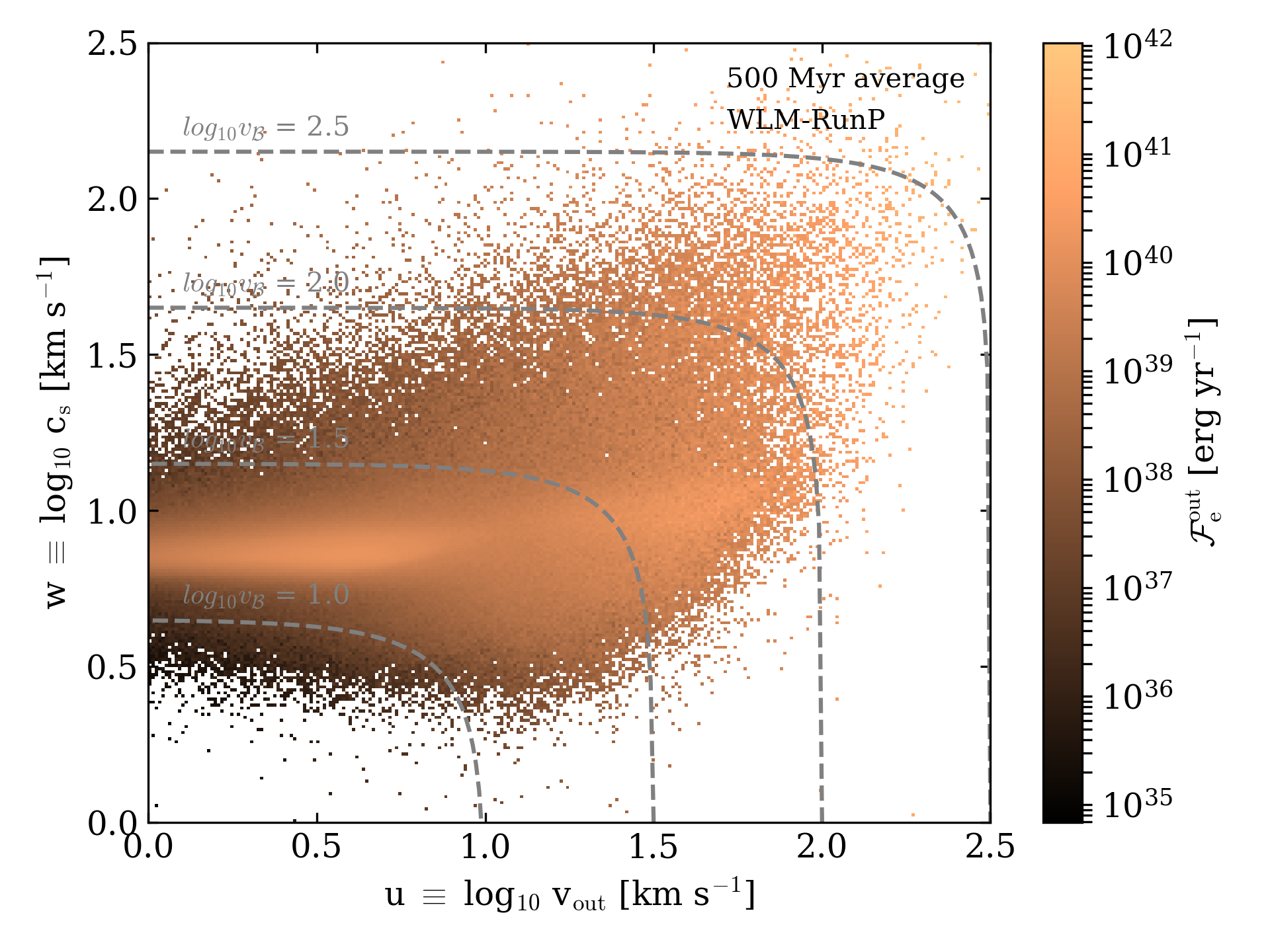}
    \caption{We show the outflow velocity as a function of the sound speed in logarithmic space for the models \textit{WLM-fiducial} (top) and \textit{WLM-RunP}. The colour code indicates the averaged outflow energy flux rate over a time span of 500 Myr. The energy flux carried by the phase at high velocities and high sound speeds is significantly increased when runaway stars are included in the run \textit{WLM-RunP} (bottom).}
    \label{fig:logu_energy}
\end{figure}

\begin{figure*}
    \centering
    \includegraphics[width=0.45\textwidth]{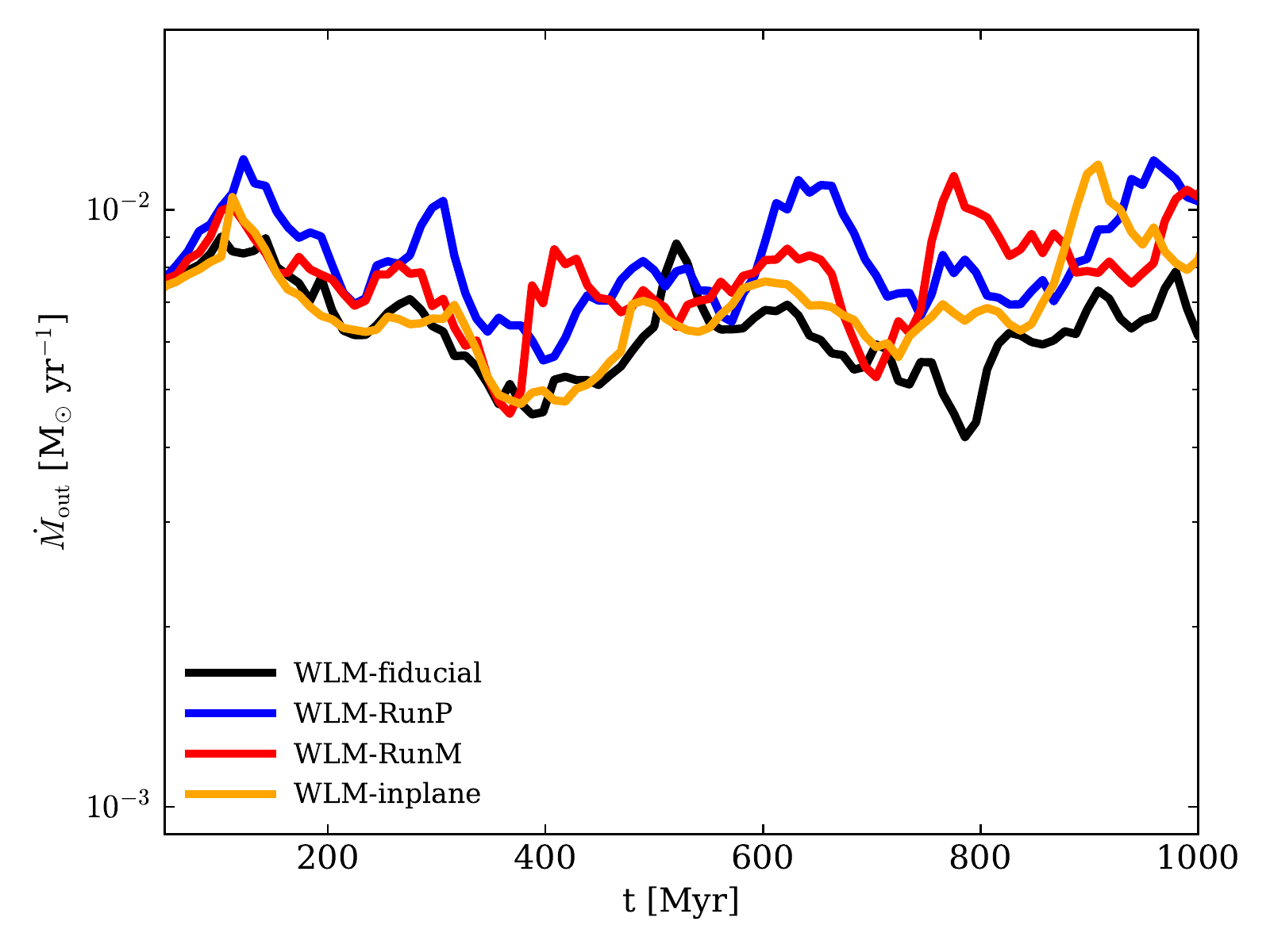}
    \includegraphics[width=0.45\textwidth]{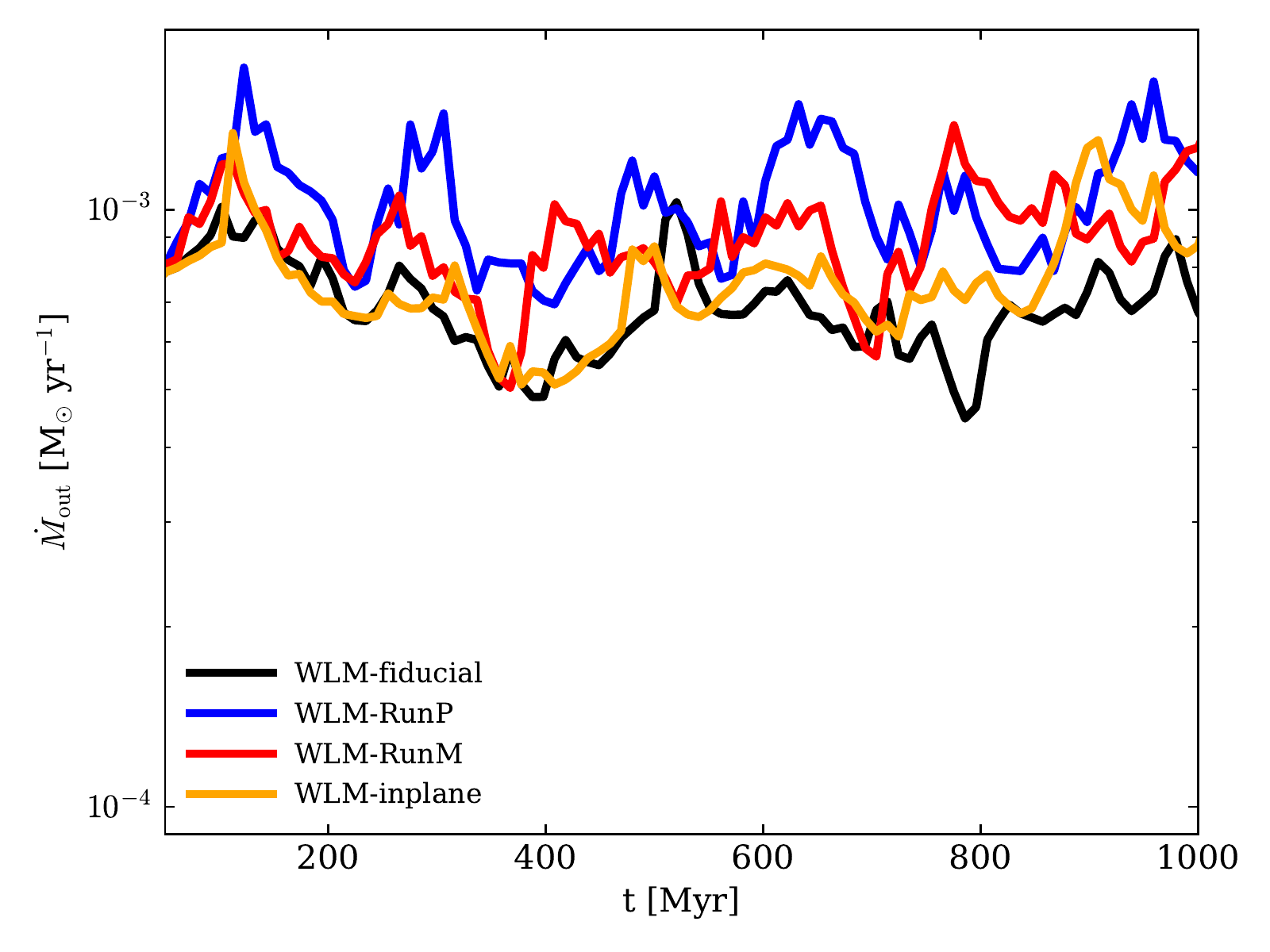}
    \includegraphics[width=0.45\textwidth]{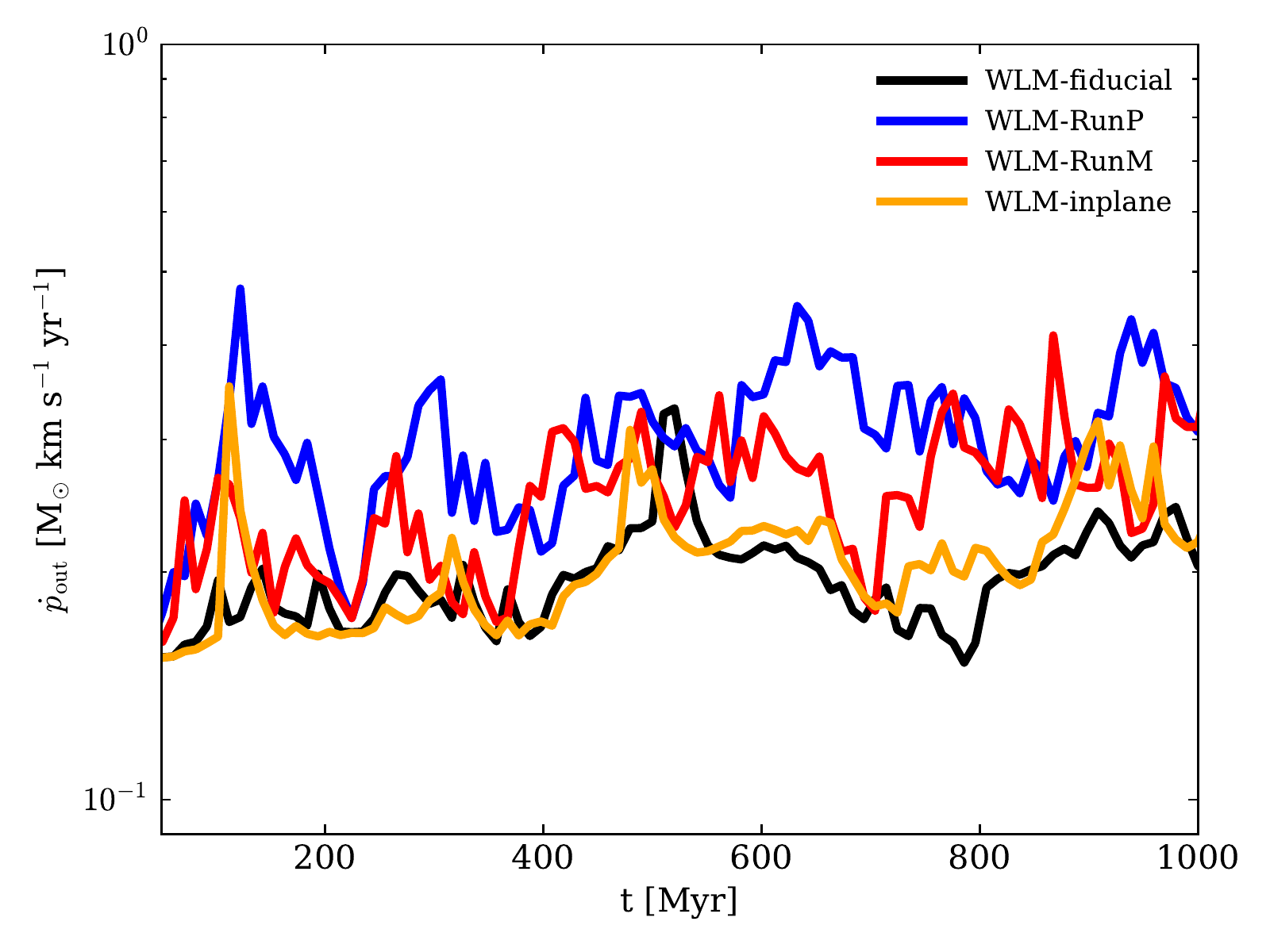}
    \includegraphics[width=0.45\textwidth]{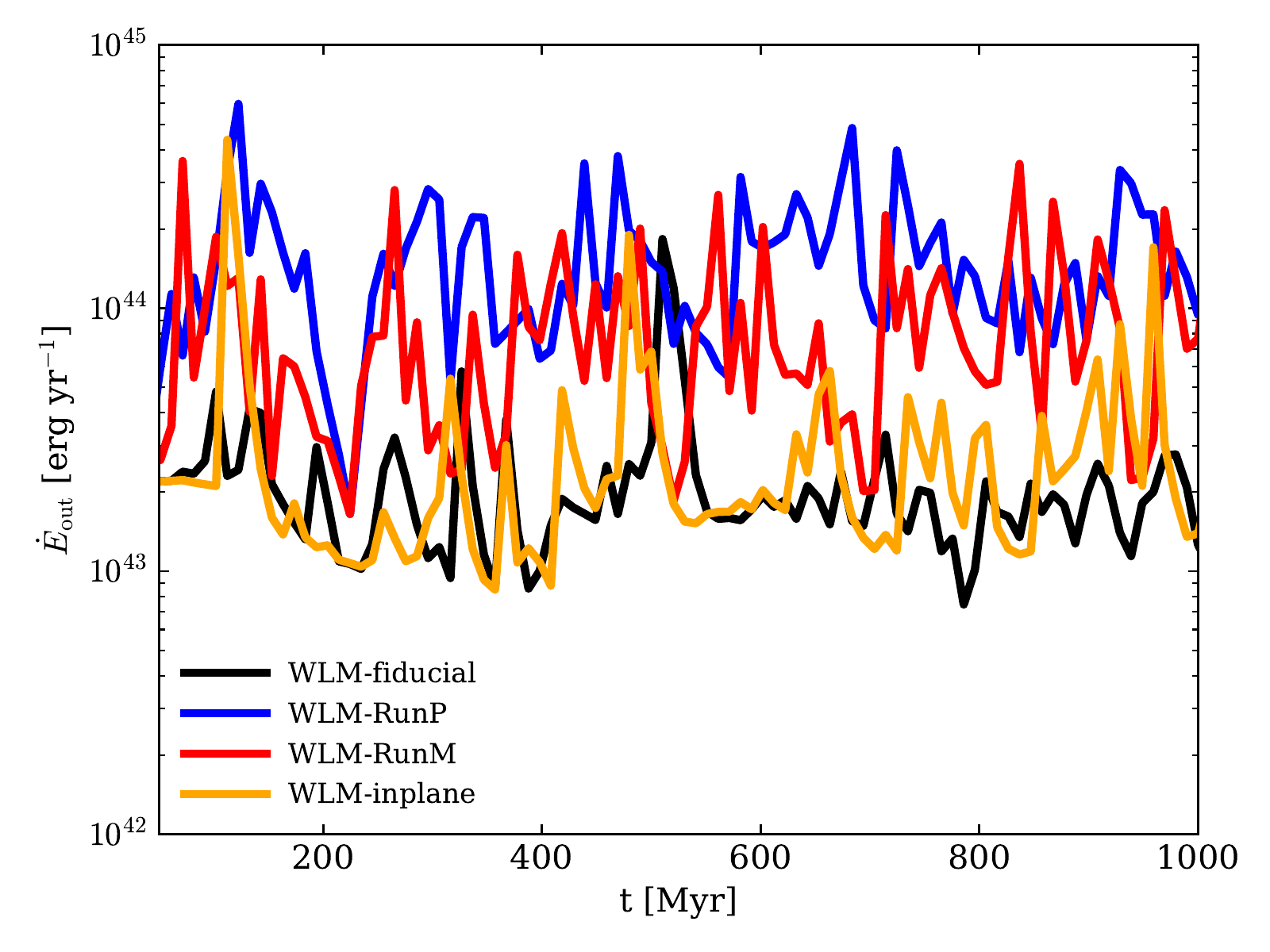}
    \caption{We show the mass outflow rate (top left), the metal outflow rate (top right), the momentum outflow rate (bottom left) and the energy outflow rate (bottom right) as a function of time for all four simulated models, where black indicates the fiducial model \textit{WLM-fiducial}, blue represents the power law model \textit{WLM-RunP }and red the Maxwellian runaway model \textit{WLM-RunM}. The orange line shows the runaway model \textit{WLM-inplane} in which we apply the velocity kicks only along the v$_\mathrm{x}$ and the v$_\mathrm{y}$ directions. The impact of runaway stars can be clearly seen in a boost of the mass, the metal and the momentum outflow rate by a factor of around two (comparing blue lines to black lines). In the case of the energy outflow rate the boost is even more significant, around a factor seven to eight.}
    \label{fig:mass_out}
\end{figure*}

\begin{figure*}
    \centering
    \includegraphics[width=0.45\textwidth]{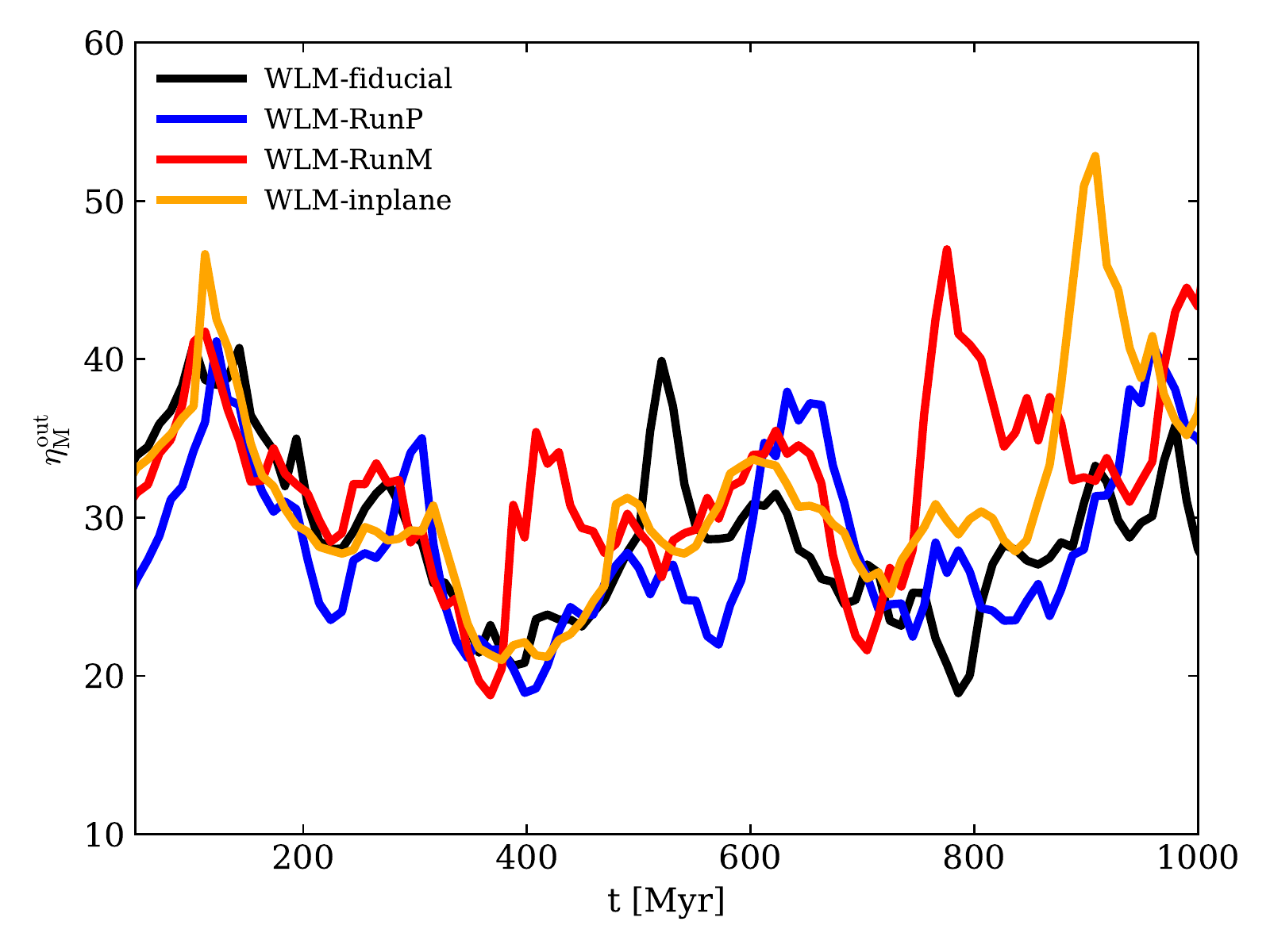}
    \includegraphics[width=0.45\textwidth]{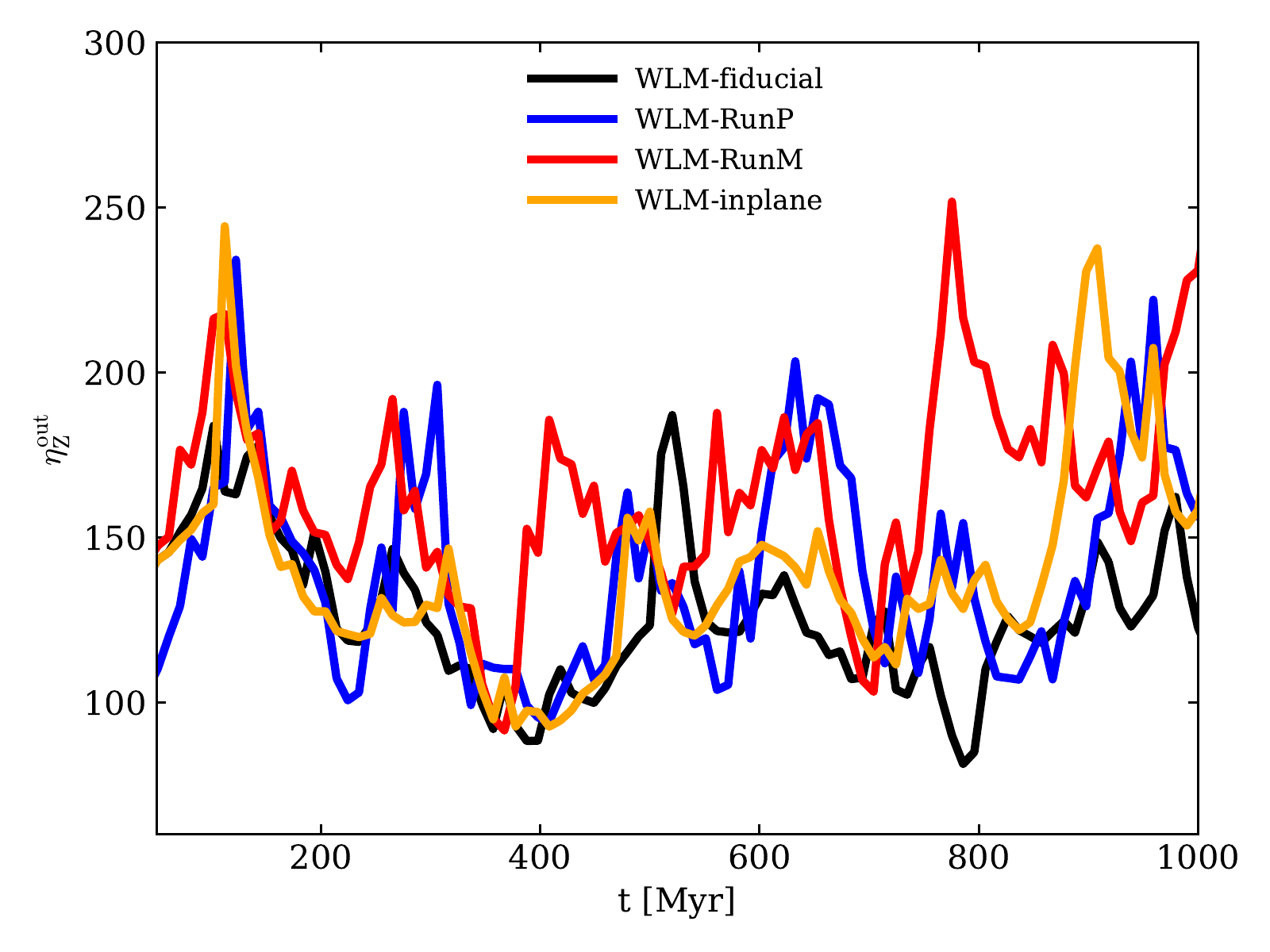}
    \includegraphics[width=0.45\textwidth]{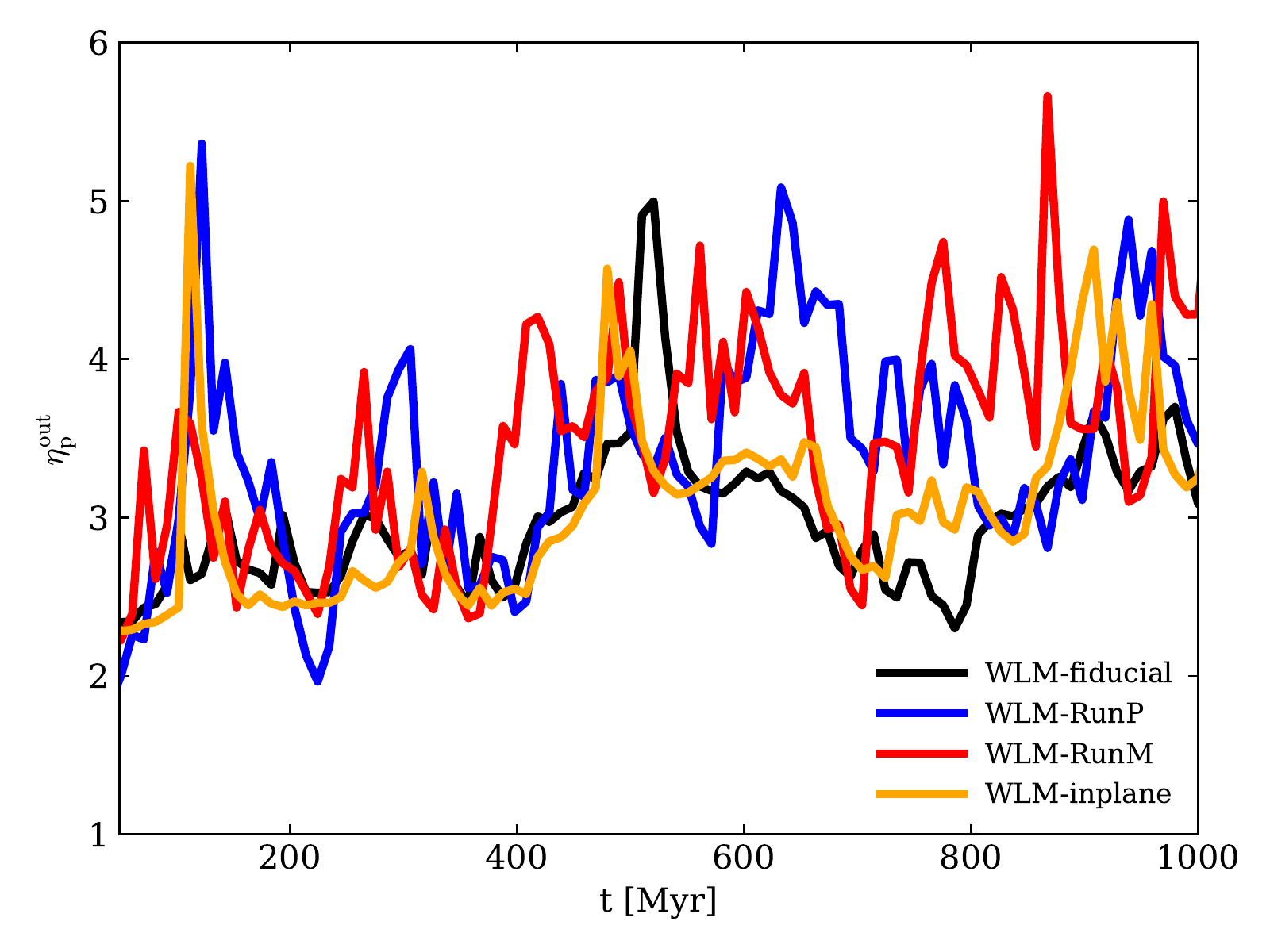}
    \includegraphics[width=0.45\textwidth]{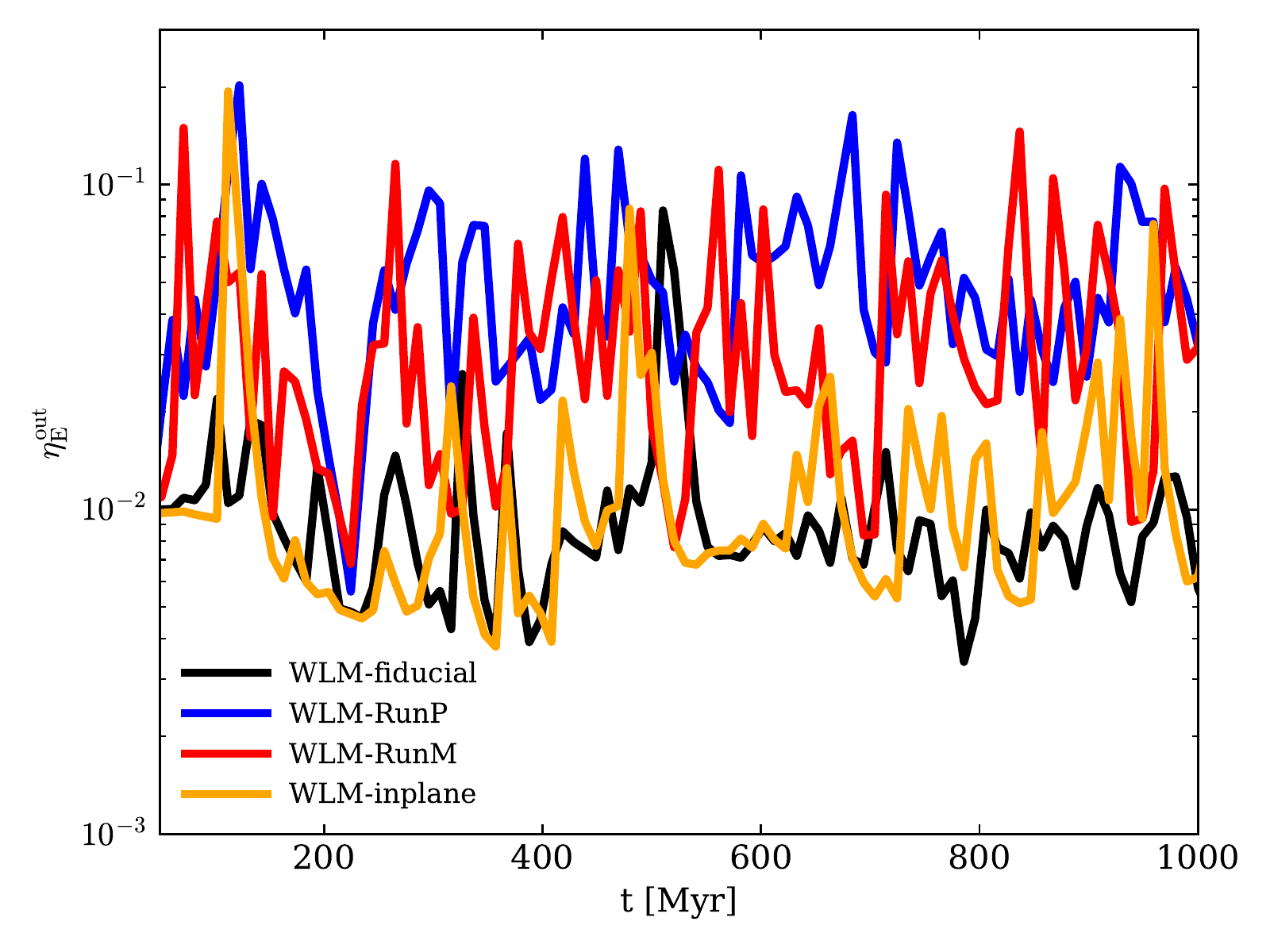}
    \caption{We show the mass outflow loading (top left), the metal outflow loading (top right), the momentum outflow loading (bottom left) and the energy outflow loading (bottom right) as a function of time for all four simulated models, where black indicates the fiducial model \textit{WLM-fiducial}, blue represents the power law model, \textit{WLM-RunP} and red the Maxwellian runaway model \textit{WLM-RunM}. The orange line shows the runaway model \textit{WLM-inplane} in which we apply the velocity kick only along the v$_\mathrm{x}$ and the v$_\mathrm{y}$ directions. While the trend of increased mass and metal outflow rates is reduced when expressed in terms of the normalized loading factors, the trend of increased momentum and energy loading remains roughly the same as that deduced from the results of Fig.~\ref{fig:mass_out}.}
    \label{fig:loading_factors}
\end{figure*}

\begin{figure*}
    \centering
    \includegraphics[width=0.45\textwidth]{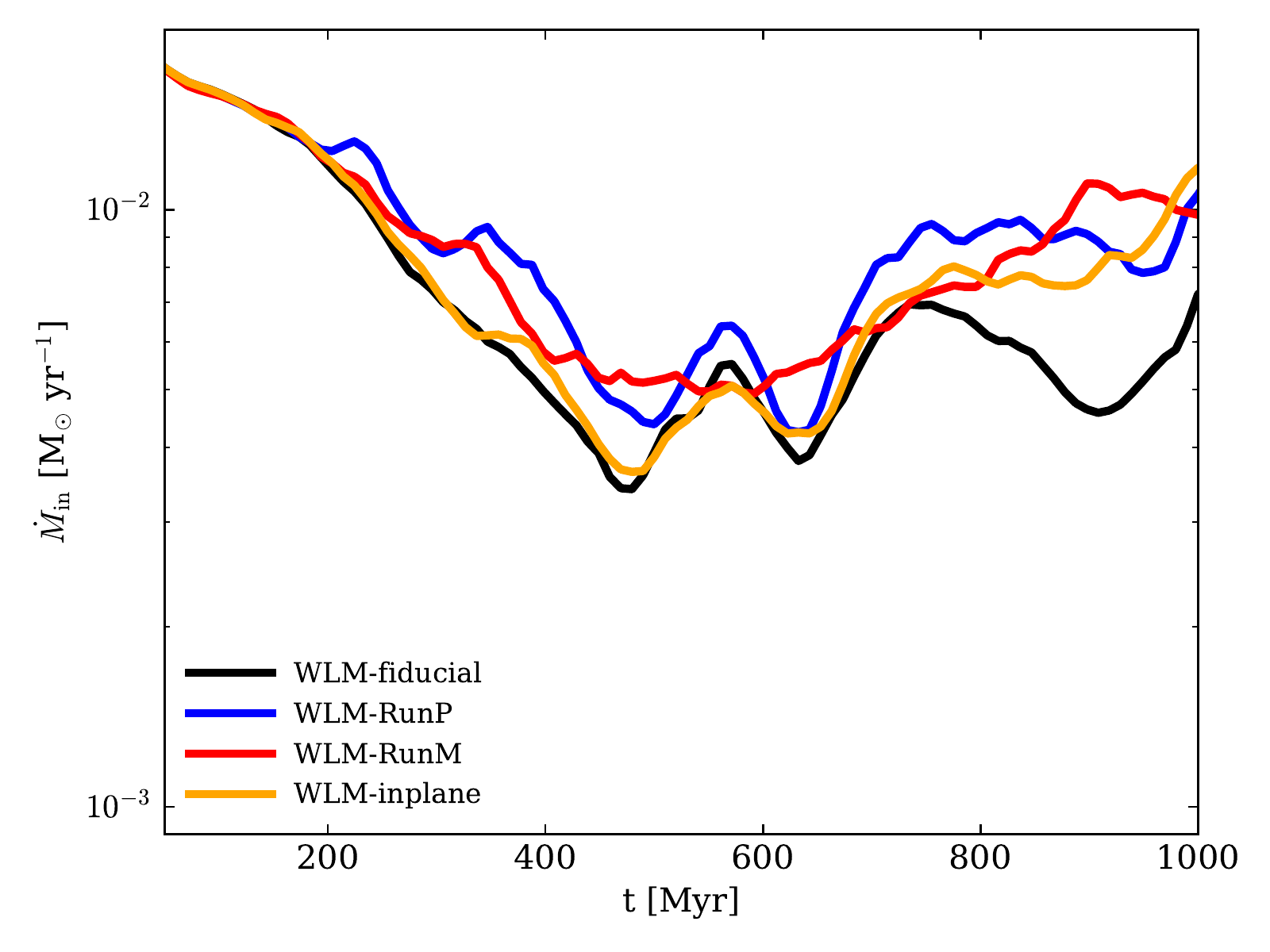}
    \includegraphics[width=0.45\textwidth]{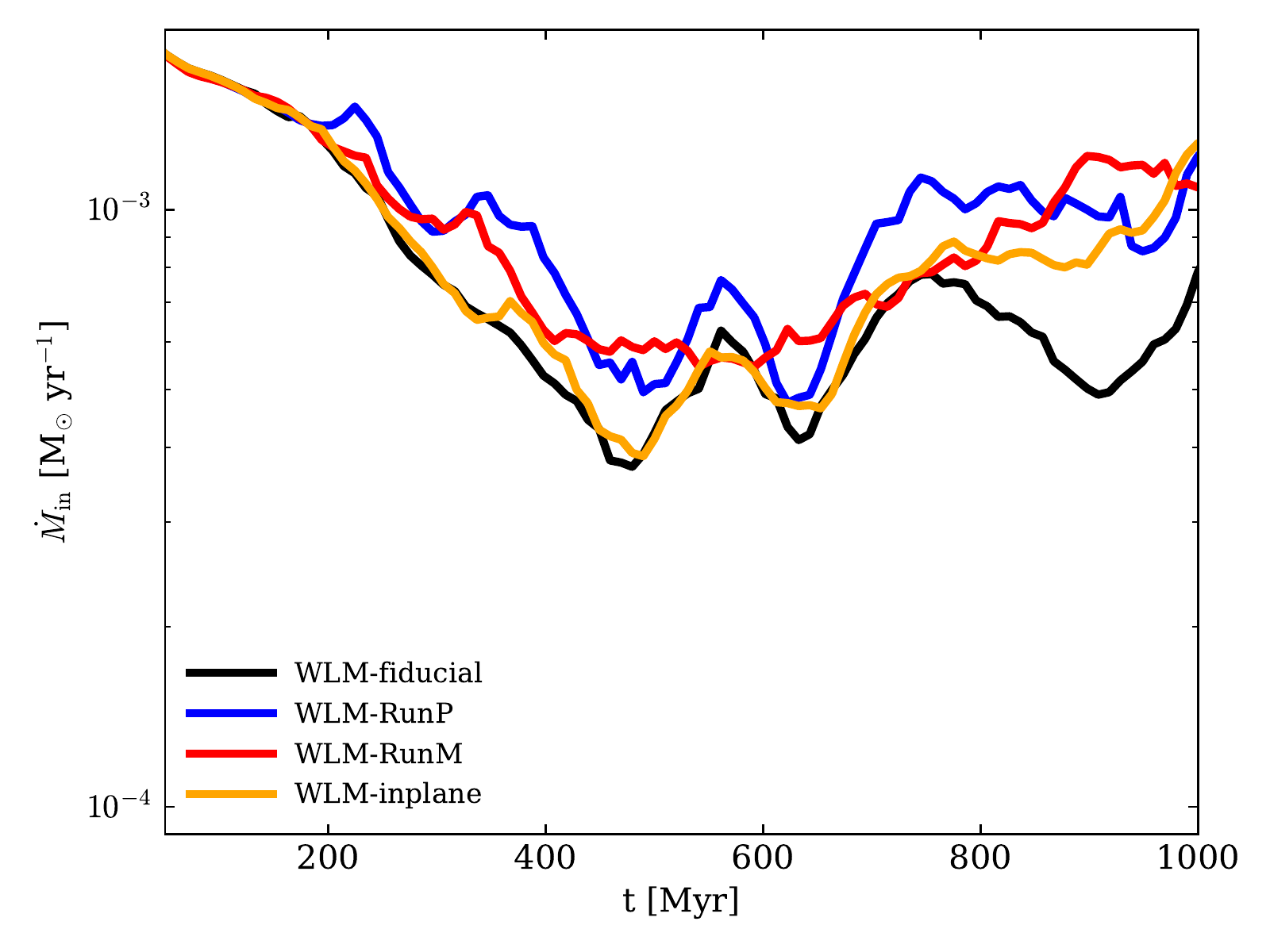}
    \includegraphics[width=0.45\textwidth]{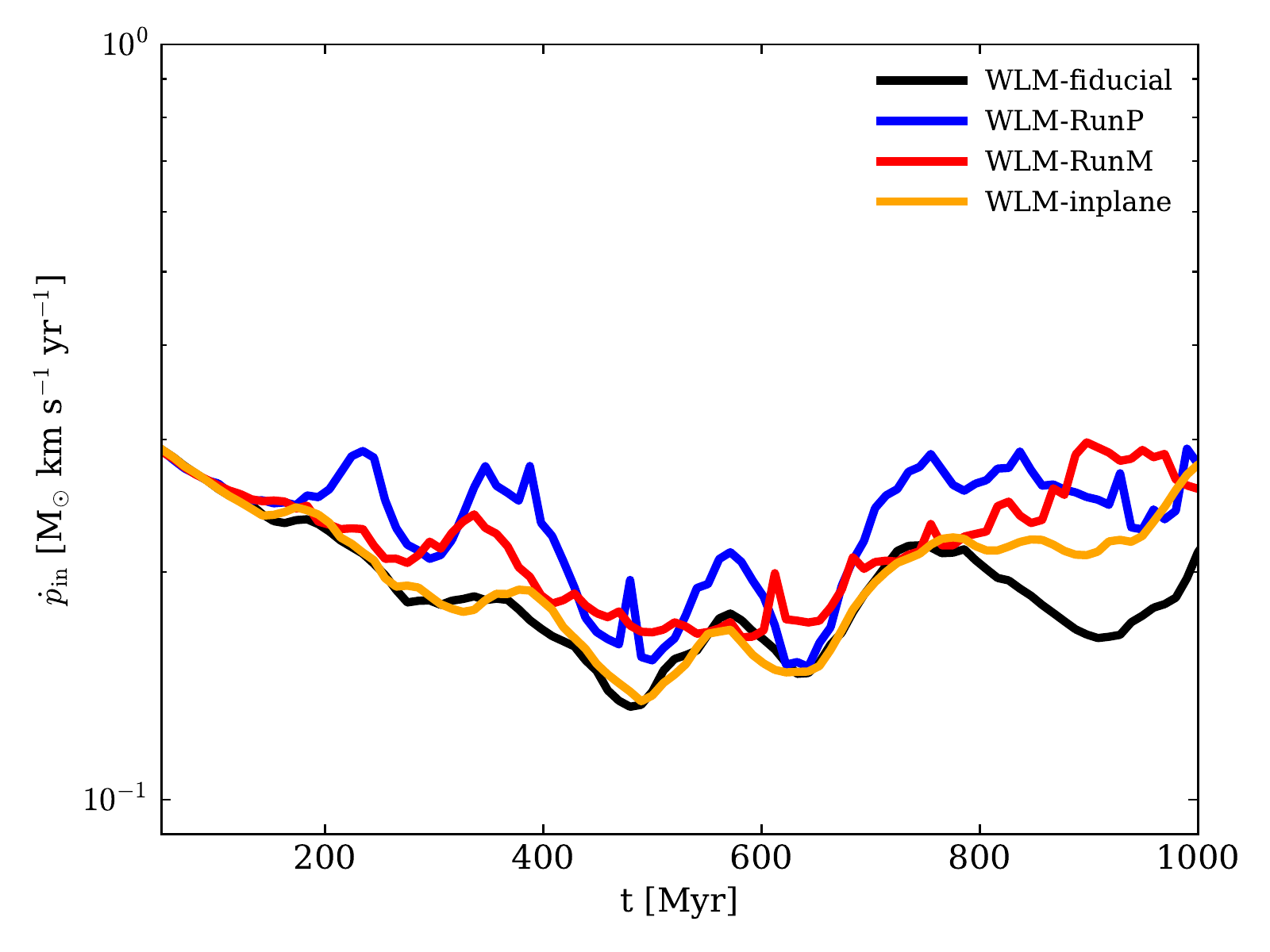}
    \includegraphics[width=0.45\textwidth]{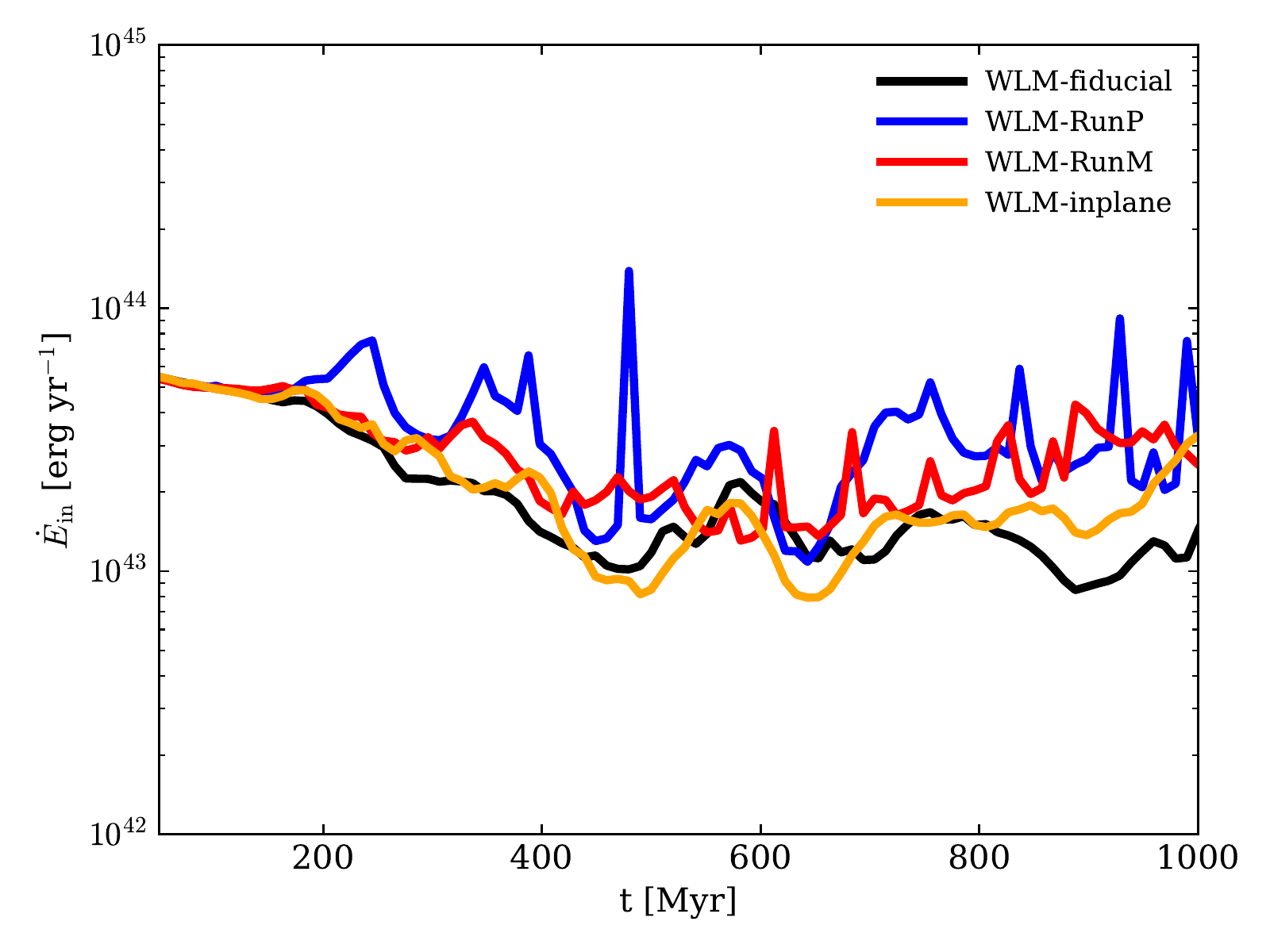}
    \caption{We show the mass inflow rate (top left), the metal inflow rate (top right), the momentum inflow rate (bottom left) and the energy inflow rate (bottom right) as a function of time for all four simulations \textit{WLM-fiducial} (black), \textit{WLM-RunP} (blue), \textit{WLM-RunM} (red) and \textit{WLM-inplane}. We find increased inflow rates for all the models that include runaway stars.}
    \label{fig:mass_in}
\end{figure*}

\subsection{Morphology}
We start the discussion of our results with a brief comparison of the morphological structure of the resolved ISM in the simulations \textit{with} and \textit{without} runaway stars. As summarized in Table \ref{tab:models} the models \textit{WLM-RunP, WLM-RunM} and \textit{WLM-inplane} include a treatment for runaway stars, while the simulation \textit{WLM-fiducial} is a control run that includes the feedback from massive stars (supernovae, photoionisation and photo electric heating), as described in Section~\ref{sec:numerical}.

In Fig.~\ref{fig:faceon} we show face-on projections of the stellar surface density (left), the gas surface density (center) and the thermal pressure (right) for the runs \textit{WLM-fiducial} (first row), \textit{WLM-RunP} (second row), \textit{WLM-RunM} (third row) and \textit{WLM-inplane} (fourth row). Qualitatively, the runs look very similar and there are only marginal differences in the global evolution of the stellar and the ISM-structure of the simulated systems. This trend is generally true over the full duration of the simulation. In the simulations that include a treatment of runaway stars, the stellar structure appears to be less clustered than in the cases without runaway stars and the regular stellar tails that are clearly visible in the run \textit{WLM-fiducial} appear to be less prominent in the simulations \textit{WLM-RunP}, \textit{WLM-RunM} and \textit{WLM-inplane}. However, the effect is marginal. \\
In Fig.~\ref{fig:edgeon} we show the same results for the simulations \textit{WLM-fiducial} (first row), \textit{WLM-RunP} (second row), \textit{WLM-RunM} (third row) and \textit{WLM-inplane} (fourth row) for the edge-on perspective. One can easily identify the thin stellar disk. 
Again we note that the overall morphology is similar in all four cases, but one can see a slighty ``puffed up'' stellar disc in the runs \textit{WLM-RunP} and \textit{WLM-RunM}, which include a treatment for runaway stars. This effect is not present in the model \textit{WLM-inplane}, in which velocity kicks are only applied in directions parallel to the galactic plane. We note that the runaway stars have a minor impact on the morphological structure of the stars and the turbulent, supernova driven ISM, relative to the variations expected just due to the stochastic nature of the underlying star formation and feedback prescription.

\subsection{Star formation rate}
In the top panel of Fig.~\ref{fig:sfr} we show the time evolution of the star formation rate (SFR) for all four simulations \textit{WLM-fiducial} (black), \textit{WLM-RunP} (blue), \textit{WLM-RunM}, and \textit{WLM-inplane} (orange). Comparable to similar simulations of the WLM system in previous work, the SFR in our fiducial run is settling around $2 \times 10^{-4}$ M$_{\odot}$ yr$^{-1}$, where we note that the star formation rate drops by roughly a factor of two due to the presence of photoionising radiation, compared to runs that only include the feedback from SNe alone \citep[e.g.][]{Hu2017, Smith2021}. Generally, the star formation rate in the two runaway models \textit{WLM-RunP} and \textit{RunM} do not differ significantly from the star formation rate in both the fiducial run \textit{WLM-fiducial} and the run \textit{WLM-inplane}, in which we only consider velocity kicks in the x and y plane, with the maxwellian velocity kick model. However, there are some differences that we would like to point out in greater detail. First, there is a difference in the strong runaway model \textit{WLM-RunP} compared to the other three runs, namely the height of the peaks in SFR which can be larger by more than a factor of two for specific selected times. Thus the overall structure of the star formation history in the strong runaway model appears to be more bursty. The most likely reason for this is that the velocity kicks that we apply in this particular case can easily drive some of the photoionising sources out of the natal clouds. Since the molecular clouds in our WLM analogues are rather small (a few times $10^3$ M$_{\odot}$ up to around a maximum of $10^{4}$ M$_{\odot}$) they are easily dispersed by the effect of photoionisation. However, if a large fraction of massive stars are quickly removed from their natal clouds,  PI-radiation will not be able to locally quench star formation as strongly as in the reference run \textit{WLM-fiducial}. This leads to a more bursty behaviour as shown in the studies of \citet{Hu2017} and \citet{Smith2021}. This is not seen to the same degree in the simulations \textit{WLM-RunM} and \textit{WLM-inplane}, which adopt the more moderate ``walkaway'' case with a Maxwellian velocity distribution, where the bulk of stars (80 per cent) receive rather small velocity kicks ($\sim10$ km s$^{-1}$) consistent with the ``walkaway'' picture.

In the bottom panel of Fig.~\ref{fig:sfr} we show the integrated star formation rate for all four runs, adopting the same colour scheme for the lines. While the build up of the total stellar mass of the system is similar, the final masses (at t=1 Gyr) are slightly different between the runs. Generally, we find that there can be a deviation of the total mass and average star formation rate of up to a few per cent due to model stochasticity. This is very apparent for the runs \textit{WLM-fiducial} and \textit{WLM-inplane} where we find agreement on the per cent level. However, in the runaway models \textit{WLM-RunM} and \textit{WLM-RunP} we find a significant increase of the total stellar mass by 35 and 85 per cent  respectively which is a significant increase compared to model scatter of around a few per cent.
We note that we use reproducible random numbers based on Gadget's random number generator for the IMF sampling routine as well as the sampling routine for the runaway stars' velocities, our simulations are prone to the ``butterfly-effect'' recently reported on in cosmological zoom simulations \citep[e.g.][]{Genel2019} and molecular cloud simulations \citep[e.g.][]{Keller2019}. Hence our results are only strictly reproducible on the same machine for the same number of MPI ranks and OPENMP threads. 
The results of the runs \textit{WLM-fiducial} and  \textit{WLM-inplane} are consistent with one another within this model stochasticity, but this is not the case for the run \textit{WLM-RunP} and \textit{WLM-RunM} which both show a significant boost in the integrated star formation history.\\
In Fig.\ref{fig:env_dens_runaways} we show the supernova environmental density for all four runs, ranging from a density of 10$^{-4}$ cm$^{-3}$ to 10$^{1}$ cm$^{-3}$. The environmental density is computed in SPH-like fashion as a kernel weighted sum over all the neighbours of a star particle is flagged as experiencing a SN. The sum is computed before the ejecta are distributed.
For all runs we plot a total of 3000 supernova explosions. However, due to the slight differences in the models there is roughly a 50 Myr range in the time where the simulations reach that value, i.e. these distributions are not plotted at the exact same time. 
The distribution of the environmental density of core collapse supernovae explosions is very similar between the different runs. There appears to be a slight increase in the number of SN explosions in low density environments in the model \textit{WLM-RunP} compared to the other three models.  Furthermore, it is interesting to point out that the distribution of densities for the control run \textit{WLM-inplane} is very similar to the other models. From this we can directly deduce that in the following, all major changes that we report on are driven by ``high altitude supernovae'', that explode in gas with similar density but above or below the plane of the disk. 

\subsection{Phase structure and volume filling fraction}

Next, we investigate the changes in the density-temperature phase space and the hot volume filling fraction of the gas. 
In Fig.~\ref{fig:phase_diagrams_avg}, we show the time averaged density-temperature phase space diagrams over a time span of 500 Myr, which is equivalent to 500 snapshots and around $5 \times 10^{9}$ particles, for the model \textit{WLM-fiducial}(top), \textit{WLM-RunP}(centre) and \textit{WLM-RunM}(bottom). A more continuously populated hot phase can be seen in both runaway models. Additionally, we note that in both runaway cases, there is significantly more mass in the diffuse hot phase around 10$^{5}$ K. Gas in HII regions is clearly apparent as a horizontal line at  10$^{4}$ K.\\
In Fig~\ref{fig:phase_diagrams_avg_vol} we show the same phase diagrams, also averaged over 500 Myrs, in a volume weighted fashion (actually we weight by effective cell size which is V$^{1/3}$), which clearly reveals the volume filling nature of the hot phase of the ISM. 

In Fig.~\ref{fig:vff}, we show the time evolution of the hot volume filling fraction for all models as a function of time.  To define the hot phase, we choose a cut of $3 \times 10^{4}$ K, which has two motivations. The first one is numerical, as this is the temperature where we transit from a non-equilibrium cooling prescription to an equilibrium cooling prescription. The second is chosen to be comparable with previous results of \citet{Hu2017}. However, we note that this differs from the definition of the hot phase in \citet{Steinwandel2020}. We compute the hot volume filling factor as the sum over the volume stored in the hot phase above $3 \times 10^4$, divided by the total volume of the simulation. The volume is computed as:
\begin{align}
    V = \frac{4}{3} \pi h_{i}^{3},
\end{align}
where h$_{i}$ is the smoothing length of particle $i$. Alternatively, one can compute the volume as $m_{i}/\rho_{i}$. We tested both methods and note that there is only a minor difference between the two, which typically remains below the per cent level.\\

We can see from Fig~\ref{fig:vff} that the volume filling fraction of hot gas in the \textit{WLM-RunP} (strong runaway) model is significantly higher than in the other models. The weaker runaway model \textit{WLM-RunM} has a slightly higher hot gas volume filling fraction than the \textit{WLM-fiducial} and \textit{WLM-inplane} models, especially at late times. We note that the exact value for the hot phase filling fraction is very sensitive to the exact value of the temperature cut, and can change by a factor of five when the cut is varied within the range of of $3 \times 10^{4}$K and $3 \times 10^{5}$K. Nevertheless, we find a consistently larger volume filling factor for the hot phase independent of which temperature cut we adopt.

Since runaway stars induce a more efficient coupling of feedback to diffuse gas, we might expect their inclusion to effect the cold dense ISM. We do note a substantial difference in the dense cold gas between our runs in Fig~\ref{fig:phase_diagrams_avg_vol}. In particular, the run \textit{WLM-RunP} has a noticeable dearth of volume-weighted gas above n$=10^3$ cm$^{-3}$. This run also has a slightly higher star formation rate overall (as seen in Fig. \ref{fig:sfr}), which suggests that the depletion time of the cold star forming gas is shorter when runaway stars are included. In this run, massive stars are more often ejected out of the natal cold dense gas into warmer and diffuse regions of the ISM. Thus, the CNM and star forming gas will be less disturbed and more likely to continue its star forming state (e.g. can continue to be Toomre/Jeans unstable). 

We now turn to a quantification of the outflow properties and loading factors. 
In Fig.~\ref{fig:logu_mass} we show the logarithm of the outflow velocity $u \equiv \log_{10} v_\mathrm{out}$ as a function of the logarithm of the sound speed $w \equiv \log_{10} c_\mathrm{s}$, where the colour shows the total outflow rate in M$_{\odot}$ yr$^{-1}$. The gray lines indicate surfaces of constant Bernoulli-velocity given via:
\begin{align}
    v_\mathcal{B} = \left(v_\mathrm{out}^2 + \frac{2 \gamma }{\gamma-1} c_\mathrm{s}^2\right)^{\frac{1}{2}},  
\end{align}
with $\gamma = 5/3$.
We note that these are identical to the two dimensional PDFs presented by \citet{Kim2020} when reduced to dimensionless units in loading factor space. Studying these diagrams for the \textit{WLM-fiducial} (top) and \textit{WLM-RunP} runs reveals that the bulk of the mass is transported in the cold phase. This is similar for both runs. However, the run \textit{WLM-RunP} reveals an extended tail in a warmer, diffuse phase. This demonstrates that runaway stars have a clear impact on the phase structure of the resulting outflows in our dwarf galaxy simulation.  \\
In Fig.~\ref{fig:logu_energy} we show the same plots for the energy outflow rate, for \textit{WLM-fiducial} on the top and \textit{WLM-RunP} on the bottom, indicating that while there is a cold energy transporting phase in our dwarf galaxy simulation, there is a tail that extends to higher specific energy. This is even more true in the bottom panel for strong runaway case \textit{WLM-RunP}, where we find a tail that not only extends to higher sound speed but also to higher outflow velocity, which is direct evidence that runaway stars can not only boost the mass flux but can additionally increase the specific energy loading as well as the total energy loading of the wind significantly.  

\subsection{Outflow rate and outflow mass loading}
We plot the main outflow diagnostics for all four simulations in Fig.~\ref{fig:mass_out}. The top left panel shows the mass outflow rate (M$_{\odot}$ yr$^{-1}$), the top right panel shows the metal outflow rate (M$_{\odot}$ yr$^{-1}$), the bottom left panel shows the momentum outflow rate (M$_{\odot}$ km s$^{-1}$ yr$^{-1}$) and the bottom left panel shows the energy outflow rate (erg yr$^{-1}$). We note that the outflow rates are measured at a height of $z=1$ kpc above the midplane in a patch of thickness 100 pc. We find a slight boost in the mass and the metal outflow rate by a factor of around two in the late evolutionary stages of the galactic disc when runaway stars are included. The effect is roughly the same between the strong runaway (\textit{WLM-RunP}) and weaker runaway (\textit{WLM-RunM}) case. The trend is slightly stronger in the metal outflow rate, which suggests that that the runaway stars deposit their ejecta directly into the outflowing gas. 
While the mass and metal outflow rates are already clearly boosted in the runs with runaway stars, the effect is even stronger for the momentum and energy outflow rates. For the momentum outflow rate we find a boost of around a factor of five, while we see a boost in the energy outflow rate by at least a factor of five and in single peaks up to a factor of around ten. Comparing with the reference runs \textit{WLM-fiducial} and \textit{WLM-inplane} clearly indicates that this boost stems from ``high-altitude supernovae'' in low density gas around the disc scale height of the gas, as this is the mean travel distance of the fast moving runaway stars. The effect is slightly suppressed in the \textit{WLM-RunM} model which is consistent with the scenario of walkaway stars. \\

In Fig.\ref{fig:loading_factors} we show the mass (top left), metal (top right), momentum (bottom left) and energy (bottom right) loading for all four simulations considered in this paper. While we see similar trends for the loading factors, we note that the trend in the mass loading is slightly suppressed compared to the boost we find in the mass outflow rate. 

\subsection{Inflow rates}
In Fig.\ref{fig:mass_in} we show the inflow rate for mass (top left), metals (top right), momentum (bottom left) and energy (bottom right) for the simulations \textit{WLM-fiducial} (black), \textit{WLM-RunP} (blue), \textit{WLM-RunM}, and \textit{WLM-inplane} (orange). We measure the inflow rate in the same 100 pc thick patch located at 1 kpc above/below the midplane. Generally, one can see that in all the models, the inflow rates decrease with time, then increase slightly after about 500 Myr. This increase is more pronounced in the models with runaway stars. In the models with runaway stars, the mass, metal and momentum inflow rates show a similar degree of increase as the increase in outflow rates. However, the energy inflow rate decreases by only a factor of 2-3, not a factor of $\sim 10$ as was seen in the energy outflow rate for the strong runaway model. This implies that the SN-feedback energy of ``high altitude supernovae'' is either efficiently kept in the CGM due to mixing with lower specific energy gas, or lost to cooling in the highly metal rich outflowing material. The latter case seems unlikely when we consider the phase-structure of the mass and volume weighted density-temperature diagrams. These reveal more mass in the diffuse 10$^5$ K regime at low densities, indicating that there is an slight increase in the virial temperature of the system when runaway stars are included. We note that inflow rates are slightly larger in the runs that include a runaway/walkaway model. This can be understood in combination with the slightly increased mass outflow rates we observe in the runaway models. We find that the peaks in outflow and inflow rate are shifted by roughly 150-200 Myr, indicating that some fraction of the gas is actually ``recylced'' over a few dynamical times of our simulated dwarf.

\subsection{Metal enrichment factor}

\begin{figure}
    \centering
    \includegraphics[scale=0.5]{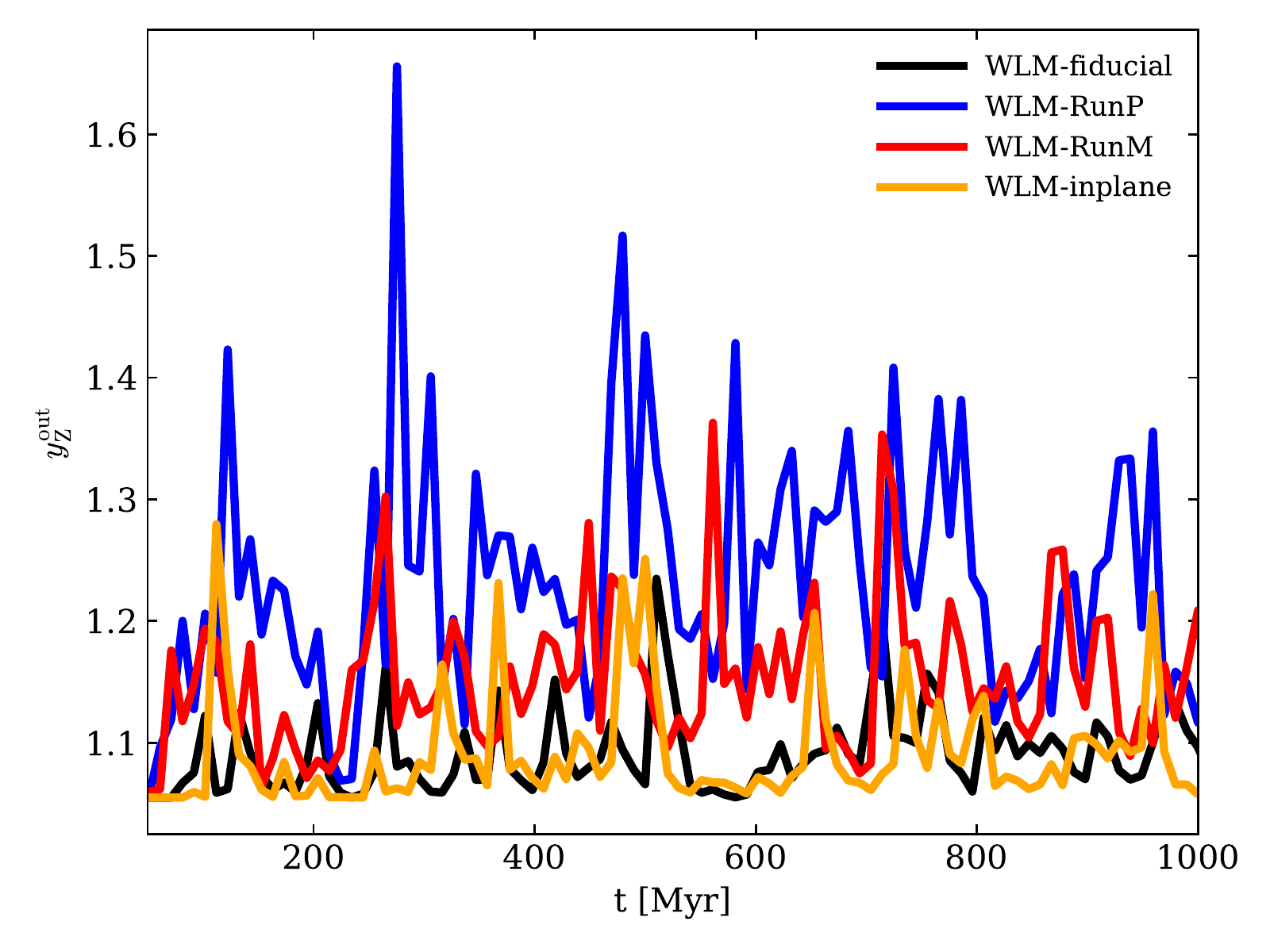}
    \includegraphics[scale=0.5]{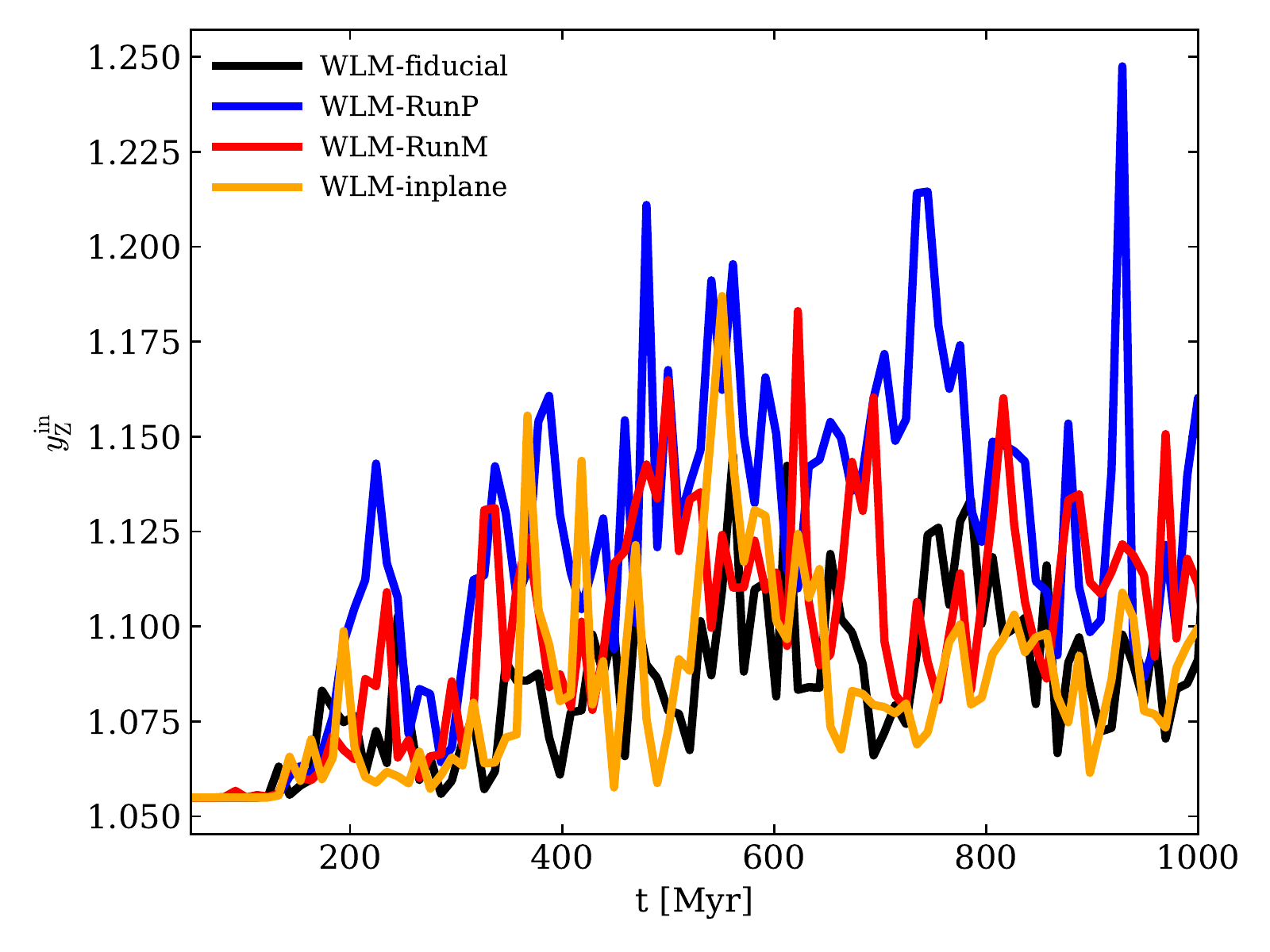}
    \caption{We show the outflow metal enrichment factor $y_{Z}^\mathrm{out}$ (top) and the inflow metal enrichment factor $y_{Z}^\mathrm{in}$ for the runs \textit{WLM-fiducial} (black), \textit{WLM-RunP} (blue), \textit{WLM-RunM} (red) and \textit{WLM-inplane} (orange). We find a significant boost in the outflow metal enrichment factor in the runs that include runaway stars. Naturally, the inflowing gas is more metal enriched as well compared to the background ISM metallicity.}
    \label{fig:metal_enrichment_factor}
\end{figure}

Finally, we discuss the metal outflow enrichment factor $y_{Z}^\mathrm{out}$ and the metal inflow enrichment factor $y_{Z}^\mathrm{in}$, which we show for all simulations in Fig.\ref{fig:metal_enrichment_factor} on the top and the bottom respectively. There is a striking increase of the metal outflow enrichment factor when runaway star modelling is included in the simulations. Generally we find a boost of around 15 to 20 percent in the metal outflow enrichment factor compared to the fiducial run and the run where we only apply the velocity kicks in the x- and y-plane. We note that while the metal inflow enrichment factor is also boosted compared to the reference simulations the effect is not as strong as for the outflow enrichment factor. This indicates that runaway stars can significantly contribute to the metal enrichment of the CGM and can boost its metallicity by about 15 percent. Since metals are the most important coolant in the low density regime, this might have crucial consequences for mixing properties in the CGM. However, since we do not explicitly model the CGM in our simulation by including a hot cooling halo, we postpone a detailed study of CGM cooling, heating and mixing rates alongside with a detailed analysis of metal emission line properties to future work. 
\section{Discussion}
\label{sec:discussion}
\subsection{Interpretation and comparison to previous work}
We start our discsussion by briefly comparing to previous results of simialr WLM systems without the inclusion of runaway star modeling, which has been used in several studies throughout the literature \citep{Hu2016, Hu2017, Emerick2019, Hu2019, Gutcke2021, Smith2021, Hislop2022}. The most appropriate of these to compare to are the simulations by \citet{Hu2017} and \citet{Smith2021} since the implemented physics modules cover the same range from the cooling and SN-feedback routines to the prescription for photo-electric heating and photoionisation. This is important since \citet{Hu2017} used a pressure-energy SPH formalism with 100 neighbours and \citet{Smith2021} adopted the moving mesh code \textsc{arepo} for their simulations which employs a second order Riemann-method on an unstaggered moving mesh. The latter is numerically more similar to our MFM approach. Generally, all codes agree quite well for our fiducial model when compared to the appropriate runs of \citet{Hu2017} (\textit{PI-SN-PE}) and \citet{Smith2021} (\textit{low$\Sigma$} in their Appendix A). This is true across star formation histories and loading factors. A more detailed study on the subtle differences between SPH and MFM on identical setups will be investigated in future work. 

Our results indicate that even conservative modelling of runaway stars in galaxy simulations with a resolved treatment of the ISM and stellar feedback can have a significant impact on the properties of large scale stellar driven winds. This most apparent in the energy outflow and energy loading factors, which we show in our respective Fig.~\ref{fig:mass_out} and Fig.~\ref{fig:loading_factors}. It is remarkable to point out that the inclusion of runaway stars can boost the energy loading factor by up to an order of magnitude, up to a value of $\sim 0.1$ in the peak values. On average, we still find an increase by roughly a factor of 5 in the model \textit{WLM-runP} and a factor of 3 in the model \textit{WLM-runM}.
We note that previous work, such as from the \textsc{tigress} suite \citep[e.g.][]{Kim2015, Kim2018, Kim2020} or from \citep[e.g.][]{Li2017}, typically find a value of the energy loading of around $\sim 0.1$ for Milky Way like conditions. 
Our results suggest that similarly high values of the energy loading might also be found in dwarf galaxies. 
The imprints of the runaway star process are not very clear in the mass and temporal weighted density-temperature phase space diagrams that we show in Fig.\ref{fig:phase_diagrams_avg}, but become more apparent in the volume weighted phase-space diagrams shown in Fig.\ref{fig:phase_diagrams_avg_vol}. There we can clearly see that the volume of the hot phase significantly increases when \textit{any} runaway modelling is included. This can be quantified not only as an increase in the specific energy loading but also in the total energy loading, as we can see in Fig.\ref{fig:logu_energy}. While the effect on the mass loading is somewhat weaker, we still see a significant increase in the mass flux rates when runaway stars are included (Fig.~\ref{fig:logu_mass}). \\

In this context it is interesting to point out that while we do not explicitly include a CGM around our dwarf galaxy simulation, it is not that case that there is no pristine low density gas around the dwarf due to the 500 Myr of initial evolution with turbulent SN-driving that we apply before examining any science results. This can be clearly deduced from Fig.~\ref{fig:env_dens_runaways} where we find little difference in the SN-environmental distribution between the fiducial run \textit{WLM-fiducial} and the run \textit{WLM-inplane} where the runaway kicks are only applied in the plane of the galaxy. The similar shape of these implies two things directly. First, the increase in the specific energy of the wind is not driven by the re-positioning of the SNe in the disc due to the presence of the runaway stars but rather stems from the ``high-altitude'' SNe, introduced by the presence of the runaway star modelling. That this effect is still present in the weaker walkaway case is a rather promising result as it indicates that the re-positioning of SNe in the vertical direction is more important at constant environmental SN-density. Second, the fact that the low density distribution of SN explosion environment is only weakly affected by the presence of \textit{any} runaway modelling indicates that the SNe have to explode around the disc scale height in gas that is still mostly resolved even without an additional CGM that would provide an external pressure. If this were not the case, the distribution of environmental SN-densities would be significantly altered and shift to an overabundance in events in unresolved low density gas. \\
Finally, it is important to point out that we find a significant increase in the metal mass loading, by around 20 per cent in both runaway models as we show in \ref{fig:metal_enrichment_factor}. This can have crucial consequences for the cooling behaviour around dwarf galaxy systems and observational properties of the CGM. \\
We can compare our results with two notable studies concerning runaway stars. One was presented by \citet{Andersson2020}, and was carried out with the adaptive mesh refinement code Ramses \citep{Teyssier2002} for a global MW-simulation. Another was carried out by \citet{Kim2018} within the \textsc{tigress} framework of local stratified environments. At face value, these studies seem to disagree with one another, as \citet{Kim2018} reports that runaways have a very weak effect, while \citet{Andersson2020} reports a significant boost in the wind outflow rates when runaway modeling is included. However, the comparison is tricky due to the different setup choices. Furthermore, both studies are evolved for a short amount of time compared to any global MW-scale. The comparison to our work is even more complicated since our simulations are designed to study very low mass dwarf galaxies at very high resolution. However, \citet{Andersson2020} state that they find that runaway stars do not have any effect in dwarf galaxies, which they discuss very briefly without showing the details of the dwarf simulations. These appear to be centred around very low mass cosmological dwarf galaxy simulations taken from the \textsc{edge} sample \citep{Agertz2020}. Indeed, we also find that if we apply the runaway modelling in lower mass isolated discs we do not see a strong effect on any of the properties, which may be because these do not form a thin disc due to the strong impact of SN-feedback. Hence our results are not necessarily in disagreement with what \citet{Andersson2020} report, but we represent difference in the galaxy properties between our study and theirs. Furthermore, we note that some of the dwarf galaxy simulations that are carried out with \textsc{ramses} by now do not see an effect of runway stars when only SN-feedback is employed (Andersson p.c. and Andersson et al. in prep.) 
The same can be said about a comparison with \citet{Kim2018}, where again, the results are very hard to compare since we are in very different outflow regime, as can be seen from the shape of our Fig.~\ref{fig:logu_mass} and Fig.~\ref{fig:logu_energy} compared to their results as shown in \citet{Kim2020}. For now we can only conclude that our results seem to be somewhere in between the studies of \citet{Andersson2020} and \citet{Kim2018}, with runaways having a weaker impact than in the Milky Way simulations of \citet{Andersson2020}, but stronger than in the work of \citet{Kim2018}. 
However, differently from both studies we find a significant increase in the appearance of high specific energy outflows although we note that while \citet{Kim2018} is investigating energy loading as well, this is not put forward in \citet{Andersson2020}. Nevertheless, we recommend that future resolved simulations include some treatment of runaways to capture this effect. However, we note that even with the runaway star modelling we do not achieve the high loading factors observed in the supernova only case \citep[e.g.][]{Hu2016, Hu2017}. The lower loading factors in out case are mostly driven by the inclusion of photoionising radiation and the runway star implementation seem to be able to counterbalance that effect on the loading factors.

\subsection{Model limitations}
In this sub-section, we briefly discuss some caveats and model limitations of our simulations. First, our two presented runaway models strictly only represent a natal kicks in a binary star formation scenario. However, we note that we only sample a single star IMF that strictly speaking does not account for a fraction of the stars to be formed in binary or few body systems. The latter would allow us to study more complex runaway scenarios like ejection of massive stars due to few body interactions or ejection of B stars in a OB binary due to supernova kicks when the O-star explodes. However, this could be realised by adding a time delay to the kick, which is straightforward to implement in our code. It is not very likely that this will have a major impact on the conclusions of our study. \\ 
Additionally, we do not include an explicit pre-existing CGM around our dwarf galaxy, which could result in an overestimate of the boost in the outflow rates and loading factors. However, there are a few arguments that support our picture even without the explicit inclusion of a low density CGM around the dwarf. First, dwarfs have a very low virial temperature of $\sim 10^{4}$ K. In combination with the assumption of a very low density in the CGM compared to the galactic disc, it is rather unlikely that the CGM of dwarf galaxies is heavily supported by thermal pressure which would work against the thermal pressure provided by high altitude SNe due to the presence of runaway stars.
Second, we note that the SN-environmental density distributions in all runs are very similar and there is no excess of SNe in lower density regions. This directly indicates that even above the disc scale height the resolution seems still to be sufficient to model the evolution of individual SN-remnants.
 
\section{Conclusions and Outlook} 
\label{sec:conclusion}

\subsection{Summary}

\begin{table*}
    \caption{Summary of the mean outflow diagnostics once the system has reached a quasi-self-regulating state} 
    \centering
    \label{tab:outflow_rate}
    \begin{tabular}{ccccccc}
      \hline
	  Name		& SFR [M$_{\odot}$ yr$^{-1}$] & $\dot{M}_\mathrm{out}$ [M$_{\odot}$ yr$^{-1}$] & $\dot{M}_\mathrm{Z, out}$ [M$_{\odot}$ yr$^{-1}$] &  $\dot{p}_\mathrm{out}$ [M$_{\odot}$ km s$^{-1}$ yr$^{-1}$] & $\dot{E}_\mathrm{out}$ [$10^{43}$ erg yr$^{-1}$]\\
	  \hline
      \textit{WLM-fiducial} & 0.00023 & 0.00632 & 0.00068 & 0.194 & 2.225 \\
      \textit{WLM-RunP} & 0.00021 & 0.00816 & 0.00100 & 0.291 & 15.298\\
      \textit{WLM-RunM} & 0.00029 & 0.00757 & 0.00087 & 0.243 & 8.512\\
      \textit{WLM-inplane} & 0.00024 & 0.00686 & 0.00074 & 0.200 & 3.085\\
      \hline
      \end{tabular}
\end{table*}

\begin{table}
    \caption{Summary of the mean loading factors once the system has reached a quasi-self-regulating state} 
    \centering
    \label{tab:loading}
    \begin{tabular}{ccccccc}
      \hline
	  Name		& $\eta_\mathrm{M}^\mathrm{out}$ & $\eta_\mathrm{Z}^\mathrm{out}$ & $\eta_\mathrm{p}^\mathrm{out}$ & $\eta_\mathrm{E}^\mathrm{out}$ & $y_\mathrm{Z}^\mathrm{out}$\\
	  \hline
      \textit{WLM-fiducial} & 31.4 & 125.2 & 2.93 & 0.11 & 1.089\\
      \textit{WLM-RunP} & 45.7 & 136.4 & 3.28 & 0.5 & 1.23\\
      \textit{WLM-RunM} & 39.0 & 143.5 & 3.34 & 0.35 & 1.15\\
      \textit{WLM-inplane} & 34.0 & 133.0 & 2.97 & 0.13 & 1.088\\
      \hline
      \end{tabular}
\end{table}

We carried out high resolution numerical simulations of isolated dwarf galaxies \textit{with} and \textit{without} a treatment for runaway stars to probe the effect of this process on galactic outflows and related galaxy and ISM properties. We showed that the effect of runaway stars is weak with respect to the resulting structure of the multiphase ISM.
Furthermore, we showed that there is an impact on the global star formation rate and the build up of the total stellar mass of the system as a function of time for all simulations that include runaway stars compared to the fiducial run. While this results in very similar distributions for the PDFs of the environmental density where SNe explode, the run \textit{WLM-RunP} shows a slight excess of supernovae in lower density environments, which is beyond the intrinsic model scatter. However, the fact that all the models, including the run \textit{WLM-inplane}, show similar SN density distributions is clear evidence that the main changes in global outflow evolution due to runaways stars is caused by the stars that explode at higher altitude around the midplane. The fact that this trend remains for the weaker runaway model \textit{WLM-RunM} indicates that runaway stars can heat the gas at the important boundary between galactic disc and CGM. \\
This can be quantified by investigating the hot phase of the ISM in density-temperature phase space in the cases with and without runaway stars. Averaging over about half of the simulation run time, we find an excess in the mass of hot gas in the diffuse low density ISM in the simulations with runaway stars. \\
In Fig.~\ref{fig:mass_out} we show the mass, metal, momentum and energy outflow rates as a function of time. All quantities are measured at a height of 1 kpc above the midplane in a thin slice with thickness 0.1 kpc. We find a slight boost in the overall mass outflow rate which is slightly more dominant when only the metal mass is considered, showing first evidence that runaway stars can lead to an excess of metals in the outflowing gas. The more dominant effect of the runaway stars can be seen in the momentum and energy outflow rates, where the momentum outflow rate is boosted by at least a factor of two in the models \textit{WLM-RunP} and \textit{WLM-RunM}. The effect is even more dramatic for the energy outflow rate where in the model \textit{WLM-RunP} it is boosted by almost one order of magnitude and in the model \textit{WLM-RunM} it increases by roughly a factor of 7. As this increase is not seen in the fiducial run \textit{WLM-fiducial} and the run \textit{WLM-inplane} in which we apply the natal kicks only in the midplane of the galaxy, this provides clear evidence that runaway stars could contribute to the driving of high specific energy winds. We note that while these trends are also seen in the momentum and energy loading they are somewhat reduced for the mass and metal outflow loading.
There is a clear trend of stronger metal enrichment in the models that include a treatment of runaway stars. We summarise the mean outflow parameters after saturation in Table \ref{tab:outflow_rate} and the mean outflow loading factors in Table \ref{tab:loading}. The key findings of this paper can be summarised as follows \begin{enumerate}
    \item Runaway stars affect the structure of the multiphase turbulent ISM and can lead to a slightly ``puffier'' stellar disc.
    \item The global star formation rate and the distribution of SN-environmental densities is increasing in the simulations with runaway stars.
    \item The cold ISM gas is less disturbed when runaways are included. This leads to enhanced star formation as the cold gas can continue to be gravitationally unstable.
    \item Runaways stars can contribute to the build-up of the hot phase of the ISM and promote outflow launching.
    \item Runaway stars can boost the mass outflow rate and metal outflow rate while keeping mass outflow and metal outflow loading roughly constant compared to the fiducial run \textit{WLM-fiducial}
    \item Runaways stars can boost the momentum outflow rate by around a factor of at least 2, a trend that remains for the momentum outflow loading.
    \item Runaways stars can boost the energy outflow rate by around a factor of 7-10 in the peaks and around 3-5 in the mean, a trend that remains for the momentum outflow loading. Therefore, they can potentially contribute to the origin of high specific energy winds in dwarf galaxies.
\end{enumerate} 

\subsection{Outlook and future work}
We close by briefly discussing a few interesting ideas and implications for future work. While we find a rather weak impact of runaway stars on the mass outflow rate and specifically the outflow mass loading, runaway stars could have implications beyond what is represented in our study. Recently, some groups have explored cosmic ray (CRs) driven winds \citep[e.g.][]{Pakmor2016, Buck2020, Hopkins2020, Hopkins2021a, Hopkins2021b, Hopkins2021c, Hopkins2021d, Hopkins2021e, Hopkins2022a, Hopkins2022b}. In particular,  \citet{Quataert2022a} and \citet{Quataert2022b} study in detail CR wind driving scenarios for CR-streaming and CR-diffusion. The outcome of this extensive analytic study is that winds by CR-streaming require a significant Alfv\'en velocity $v_\mathrm{A}$ and thus magnetic field strength. This is intrinsically hard to obtain in dwarf galaxies as the ISM turbulence is rather weak due to low SN-rates, which limits small-scale turbulent dynamo action. Furthermore, since dwarfs are often only weakly rotationally supported, there is no large scale dynamo possible. 
On the other hand, the diffusion driven winds require a large diffusion coefficient in order to obtain consistency with gamma-ray observations from pion decays that originate from cosmic ray protons, which puts rather tight constraints on the exact value of this coefficient. An interesting direction to explore in future work (both numerically and analytically) would be to constrain the effect of cosmic rays seeded by runaway stars at or above the wind launching scale which could significantly contribute to outflows driven by cosmic rays above the midplane. Hence, one could imagine runaways stars as an external boost for the cosmic ray diffusion coefficient. This could be explored with future simulations and improved analytic modelling.  

\section*{Data Availability Statement}

The data used in this article will be made available based on reasonable request to the corresponding author.

\section*{Acknowledgements}
UPS is supported by a Flatiron Research Fellowship (FRF) at the Center of Computational Astrophysics at the Flatiron Institute. The Flatiron Institute is supported by the Simons Foundation. RSS acknowledges support from the Simons Foundation.  B.B.
is grateful for funding support from the Simons Foundation, Sloan Foundation, and the Packard Foundation.
UPS acknowledges the useful discussions with Eric Andersson, Drummond Fielding, Chang-Goo Kim, Thorsten Naab, Eve Ostriker and Romain Teyssier. UPS acknowledges the computing time provided by the Leibniz-Rechenzentrum (LRZ) in Garching on SuperMUC-NG under the project number pn72bu, as well as the computing time on the c2pap-cluster in Garching under the project number pr27mi on which most the simulations presented here have been carried out. 




\bibliographystyle{mnras}
\bibliography{paper} 


\bsp	
\label{lastpage}
\end{document}